\documentclass[11pt,a4paper]{article}

\usepackage{fullpage}

\usepackage{algorithm}
\usepackage{algpseudocode}
\algtext*{EndWhile}{}
\algtext*{EndFor}{}
\algtext*{EndIf}{}

\usepackage{amssymb, amsmath, amsthm}
\usepackage{thmtools}
\usepackage{thm-restate}

\usepackage[colorlinks,linkcolor=blue,filecolor=blue,citecolor=blue,urlcolor=blue,pagebackref]{hyperref}

\usepackage{xcolor}
\usepackage{graphicx}
\graphicspath{{figs/}}

\allowdisplaybreaks

\usepackage{enumerate}
\usepackage{multicol}

\declaretheorem[name=Theorem,numberwithin=section]{theorem}

\newtheorem{corollary}[theorem]{Corollary}
\newtheorem{definition}[theorem]{Definition}
\newtheorem{lemma}[theorem]{Lemma}
\newtheorem{observation}[theorem]{Observation}

\newcommand{\av}{\mathtt{av}}

\newcommand{\R}{\mathbb{R}}
\newcommand{\cost}{\mathtt{cost}}

\newcommand{\calS}{{\mathcal{S}}}	

\newcommand{\Z}{\mathbb{Z}}
\newcommand{\rmT}{{\mathrm{T}}}
\newcommand{\bfT}{{\mathbf{T}}}

\newcommand{\ceil}[1]{\left\lceil #1\right\rceil}
\newcommand{\floor}[1]{\left\lfloor #1\right\rfloor}
\newcommand{\poly}{\mathrm{poly}}
\newcommand{\opt}{\mathrm{opt}}

\DeclareMathOperator{\E}{\mathbb{E}}
\DeclareMathOperator{\union}{\bigcup}

\newcommand{\lp}{\mathtt{lp}}
\newcommand{\ball}{\mathtt{ball}}

\mathchardef\mhyphen="2D
\newcommand{\topnorm}[1]{{\mathtt{top\mhyphen}#1}}
\newcommand{\topcn}{{\topnorm{cn}}}
\newcommand{\topell}{{\topnorm{\ell}}}

\title{On Tight FPT Time Approximation Algorithms for $k$-Clustering Problems}
\date{}
\author{
Han Dai \footnote{The work of HD and SL is supported by State Key Laboratory for Novel Software Technology, and New Cornerstone Science Foundation.} \\School of Computer Science,\\ Nanjing University, \\ Nanjing, Jiangsu Province, China. \\ \href{mailto:handai@smail.nju.edu.cn}{handai@smail.nju.edu.cn}
 \and
Shi Li \footnotemark[\value{footnote}] \\
School of Computer Science,\\ Nanjing University, \\ Nanjing, Jiangsu Province, China. \\ \href{mailto:shili@nju.edu.cn}{shili@nju.edu.cn}
\and
Sijin Peng \\
CSAIL,\\ Massachusetts Institute of Technology, \\ Cambridge, MA, United States. \\ \href{mailto:sijinp@mit.edu}{sijinp@mit.edu}
}
 
\begin{document}
    \maketitle
     \begin{abstract}
   Following recent advances in combining approximation algorithms with fixed-parameter tractability (FPT), we study FPT-time approximation algorithms for minimum-norm $k$-clustering problems, parameterized by the number $k$ of open facilities.

For the capacitated setting, we give a tight $(3+\epsilon)$-approximation for the general-norm capacitated $k$-clustering problem in FPT-time parameterized by $k$ and $\epsilon$. Prior to our work, such a result was only known for the capacitated $k$-median problem \cite{cohen2022fixed}.  As a special case, our result yields an FPT-time $3$-approximation for capacitated $k$-center. The problem has not been studied in the FPT-time setting, with the previous best known polynomial-time approximation ratio being 9 \cite{an2015centrality}.

In the uncapacitated setting, we consider the $\topcn$ norm $k$-clustering problem, where the goal of the problem is to minimize the $\topcn$ norm of the connection distance vector. Our main result is a tight $\big(1 + \frac 2{ec} + \epsilon\big)$-approximation algorithm for the problem with $c \in \big(\frac1e, 1\big]$. (For the case $c \leq \frac1e$, there is a simple tight $(3+\epsilon)$-approximation.) Our framework can be easily extended to give a tight $\left(3, 1+\frac2e + \epsilon\right)$-bicriteria approximation for the ($k$-center, $k$-median) problem in FPT time, improving the previous best polynomial-time $(4, 8)$ guarantee \cite{Alamdari2017}.

All results are based on a unified framework: computing a $(1+\epsilon)$-approximate solution using $O\left(\frac{k\log n}{\epsilon}\right)$ facilities $S$ via LP rounding, sampling a few client representatives $R$ based on the solution $S$, guessing a few pivots from $S \cup R$ and some radius information on the pivots, and solving the problem using the guesses. We believe this framework can lead to further results on $k$-clustering problems. 
\end{abstract}
    \section{Introduction}

In the supplier setting of clustering problems, we are given a set $F$ of facilities, a set $C$ of $n$ clients, a metric $d$ over $F\cup C$, and a non-negative integer $k$.  The objective is to open a set $S \subseteq F$ of $k$ facilities and assign each client to an open facility, so as to minimize some function on the connection distances.  Chakrabarty and Swamy~\cite{Chakrabarty2018} introduced the \emph{minimum-norm $k$-clustering problem}, where the goal is to minimize a monotone symmetric norm $f: \R_{\geq 0}^C \to \R_{\geq 0}$ applied to the connection distances. Specifically, the objective is to find a set $S \subseteq F$ of $k$ facilities so as to minimize $f((d(j, S))_{j \in C})$, where $d(j, S):=\min_{i \in S}d(j, i)$. 

When the norm $f$ is $L_\infty, L_1$ or $L_2$ norm, the problem respectively corresponds to the well-known \emph{$k$-supplier-center}, \emph{$k$-median} and \emph{metric $k$-means}\footnote{The objective of the metric $k$-means problem is to minimize the sum of squared distances, which is equivalent to minimize the $L_2$ norm.} problems. These problems have been extensively studied in the literature \cite{Ahmadian2020,  Charikar1999, Byrka2017,  hochbaum1985best, gonzalez1985clustering, cohen2021near, feder1988optimal, eisenstat2014approximating, conf/focs/JainV99, cohen2022improvededuclidean, feldman2007ptas, friggstad2019local, lattanzi2019better}.
The best known approximation ratios for the three problems are respectively $3$ \cite{Plesnik1987}, $2+\epsilon$ \cite{cohen20252+} and $6.357$ \cite{Ahmadian2020}, while the known hardness of approximation bounds are $3, 1 + \frac2e$ and $1+\frac8e$ \cite{hochbaum1986unified, guha1999greedy}.

Two other important classes of monotone symmetric norms studied in the literature are the $\topell$ norms and \emph{ordered} norms. For any $\ell \in [n]$, the $\topell$ norm of $v \in \R_{\geq 0}^n$ is the sum of $\ell$ largest coordinates of $v$. An ordered norm is a non-negative linear combination of several $\topell$ norms. The clustering problem with this objective is called the \emph{ordered $k$-median} problem. Independently, Byrka,  Sornat, and Spoerhase \cite{Byrka2018} and Chakrabarty and Swamy \cite{Chakrabarty2018} gave the first constant-factor approximation algorithms for ordered $k$-median. The approximation ratio was subsequently improved by Chakrabarty and Swamy \cite{Chakrabarty2019} to $5 + \epsilon$. \smallskip

Another aspect that naturally arises in clustering problems is the presence of facility capacities. In the supplier setting, a facility may not be able to serve too many clients. In clustering applications, there may be upper bounds on the sizes of clusters. In the \emph{minimum-norm capacitated $k$-clustering} problem, each facility $i \in F$ is additionally given a capacity $u_i \in \Z_{>0}$. In a feasible solution, every facility $i \in S$ can serve at most $u_i$ clients. Formally, we need to output a set $S \subseteq F$ of $k$ facilities and an assignment vector $\sigma \in S^C$ such that $|\sigma^{-1}(i)| \leq u_i$ for every $i \in S$, so as to minimize $f(d(j, \sigma_j)_{j \in C})$. 

The main problems that have been studied in the capacitated setting are the capacitated $k$-center and capacitated $k$-median problems. For capacitated $k$-center, the current best approximation ratio is $9$, due to \cite{an2015centrality}, which improves upon the earlier $O(1)$-approximation of Cygan, Hajiaghayi and Khuller \cite{cygan2012lp}. When capacities are uniform, a better approximation ratio of $6$ is known \cite{barilan1993allocate, khuller2000capacitated}. In contrast, the approximation status of the capacitated $k$-median problem is less satisfactory. Many $O(1)$ bicriteria approximation algorithms are known, which either violate the limit $k$ on the number of open facilities~\cite{aardal2015approximation} or the capacity constraints~\cite{chuzhoy2005approximating, li2014improved, byrka2014bi}. In both settings, the violation factor can be reduced to $1 + \epsilon$~\cite{Li2017uniform, Li2016, byrka2016approximation, demirci2016constant}. 
For true approximation algorithms, the folklore result of $O(\log n)$-distortion embedding of any $n$-point metric into a distribution of HST metrics   \cite{fakcharoenphol2003tight} leads to an $O(\log n)$-approximation for the problem. This ratio was improved to $O(\log k)$ by Adamczyk, Byrka, Marcinkowski, Meesum and Wlodarczyk \cite{Adamczyk2019}, who applied the HST embedding technique, but on a $k$-point metric obtained by considering the $k$-center objective. 
This remains the  current best approximation ratio for the capacitated $k$-median problem, and obtaining a true $O(1)$-approximation for this problem---without any violation---remains a notorious open question.   \smallskip

Recently, combining paradigms of approximation algorithms and \emph{Fixed-Parameter-Tractability (FPT)} has led to exciting new results, especially for problems where the improvement on polynomial time approximation algorithms has stalled, including many clustering problems.   In many clustering applications, $k$ is a small number, allowing us to have a running time of the form $g(k) \cdot n^{O(1)}$, where $g$ is a function only of $k$. Such a running time is called FPT time, parameterized by $k$. 

Cohen-Addad, Gupta, Kumar,  Lee and Li \cite{DBLP:Cohenaddad2019} studied the $k$-median and metric $k$-means problems in FPT time, obtaining approximation factors of $1 + \frac{2}{e} + \epsilon$ and $1 + \frac{8}{e} + \epsilon$ respectively. On the negative side, these approximation factors are tight in FPT time under the assumption $\text{FPT} \neq W[1]$ \cite{DBLP:Cohenaddad2019}. For the capacitated $k$-median problem, Cohen-Addad and Li \cite{cohen2022fixed} developed a tight $3$-approximation in FPT time. More recently, Abbasi, Banerjee, Byrka, Chalermsook, Gadekar,
Khodamoradi, Marx, Sharma, and Spoerhase \cite{Abbasi2023} proposed an Efficient Parameterized Approximation Scheme (EPAS) for the minimum-norm $k$-clustering problem on metrics with bounded $\epsilon$-scatter dimension, which include Euclidean metrics, metrics of bounded doubling dimension and planar metrics. This scheme achieves a $(1+\epsilon)$-approximation in $g(k, \epsilon)\cdot \poly(n)$ time, for some function $g$ on $k$ and $\epsilon$.

Goyal and Jaiswal \cite{tcs/GoyalJ23} studied FPT time approximation algorithms for a family of constrained clustering problems, with the objective of minimizing the $z$-th power of the maximum connection distance for any $z > 0$. A main constraint they consider is the \emph{cluster-size constraint}: we are given $k$ size bounds $r_1, r_2, \cdots, r_k \geq 0$, and the partition $(O_1, O_2, \cdots, O_k)$ of clients must satisfy $|O_1| \leq r_1, |O_2| \leq r_2, \cdots, |O_k| \leq r_k$ (the partition can be arbitrarily permuted). They also considered many other constraints such as lower bounds on cluster sizes, color constraints, and fault-tolerant requirements. They developed a general FPT-time framework for these problems that achieve an approximation ratio of $2^z$ for $k$-center-type problems (where facilities can be put anywhere in the metric), and $3^z$ for $k$-supplier-type problems (where facilities can only be put at specified locations). Both ratios were shown to be tight under GAP-ETH.  

In particular, the problem most closely related to the capacitated $k$-center problem we study is the $r$-capacity $k$-supplier problem. The goal is to find a clustering of clients satisfying the cluster-size constraint and choose a facility for each cluster, so as to minimize the maximum connection distance. When all size bounds $r_1, r_2, \cdots, r_k$ are the same, the problem reduces to the soft and uniform capacitated $k$-center problem where each facility $i$ has capacity $u_i = r_1$. However, as noted in \cite{tcs/GoyalJ23}, for general (non-uniform) size bounds, there is no easy reduction between the two problems, since one imposes capacities on clusters whereas the other imposes capacities on facilities.

\subsection{Our Results} 
Following this line of research, we study approximation algorithms for $k$-clustering problems in FPT time, with $k$ being the parameter. 

Our first result is a $(3+\epsilon)$-approximation algorithm for the minimum-norm capacitated $k$-clustering problem in FPT time parameterized by $k$ and $\epsilon$. 
\begin{theorem}
    \label{thm:topell-kC}
    For any $\epsilon > 0$, there is a $(3+\epsilon)$-approximation algorithm for the minimum-norm capacitated $k$-clustering problem, that runs in time $g(k, \epsilon)\cdot \poly(n)$, where $g$ is a computable function depending  on $k$ and $\epsilon$. 
\end{theorem}
                                   
In particular, this implies
\begin{corollary}
    There is a $3$-approximation algorithm for the capacitated $k$-center problem, that runs in time $g(k)\cdot \poly(n)$, where $g$ is a computable function depending on $k$. \footnote{For the $k$-center objective, $3+\epsilon$ can be reduced to $3$. By guessing and scaling distances, we can assume the optimum value is $1$. Changing the distance $d(i, j)$ to $\ceil{d(i, j)}$ does not change the optimum solution of the instance. For such an instance, a $(3 + \epsilon)$-approximation is a $3$-approximation if $\epsilon < 1$.}
\end{corollary}

To the best of our knowledge, the result was previously only known for capacitated $k$-median \cite{cohen2022fixed}. In particular, it was not known even for the capacitated $k$-center problem with general capacities. The best known approximation ratio for the problem remains $9$, achieved by a polynomial time algorithm due to \cite{an2015centrality}. Our result gives a $3$-approximation in FPT time. This is tight under the assumption that $\text{FPT}\neq W[1]$, even for the (uncapacitated) $k$-supplier center problem. \medskip

We then turn to the uncapacitated setting. Our main focus is the $\topcn$ norm for a constant $c \in (0, 1]$. (The problem can be refered to as the $cn$-centrum problem in the literature.) On the negative side, the problem is hard to approximate within a factor better than $\min\{3, 1 + \frac{2}{ec}\}$ in FPT time, assuming $\text{FPT}\neq W[1]$ \cite{osadnik2023fixed}.  We complement this with a matching positive result: 
\begin{theorem}\label{theo: top-cn norm}
     For any $\epsilon > 0$, there is a $\big(\min\big\{3, 1+\frac{2}{ec}\big\} + \epsilon\big)$-approximation algorithm for the $\topcn$ norm $k$-clustering problem with running time $\left(\frac{k}{\epsilon}\right)^{O(k)}\cdot n^{O(1)}$.%, where $g$ is a computable function on $k$ and $\epsilon$.
\end{theorem}
The theorem suggests that considering the $\topell$ norm for $\ell = cn$ is appropriate, as the approximation ratio is a function of $\frac{\ell}{n}$. The negative result implies that if $\frac{\ell}{n} < \frac1e + \epsilon$, then the problem is hard to approximate within a factor better than $3$ even in FPT time, assuming $\text{FPT} \neq W[1]$. On the other hand, achieving a $(3 + \epsilon)$-approximation for minimum-norm $k$-clustering in FPT time is easy (See Corollary~\ref{coro:three-approximation}).
\smallskip

Finally, we show that the framework can be easily extended to give a $(3, 1 + \frac2e + \epsilon)$-bi-criteria approximation for the clustering problem with both $k$-center and $k$-median objectives. This was introduced by Alamdari and Shmoys~\cite{Alamdari2017}, who gave a polynomial-time $(4, 8)$-approximation. We improve the bi-criteria approximation factors to $(3, 1 + \frac2e + \epsilon)$, albeit with FPT time. Indeed, our result works for the more general $\topcn$ norm. See Section~\ref{sec:2-norms} for formal definitions used in the theorem:
\begin{theorem}\label{thm:bi-criteria}
    For a constant $\epsilon > 0$, there is a $\Big(3, \min\big\{3, 1+\frac{2}{ec}\big\} + \epsilon\Big)$-bi-criteria approximation algorithm for the $(L_\infty, \topcn)$-norms $k$-clustering problem, with running time $\left(\frac{k}{\epsilon}\right)^{O(k)}\cdot \poly(n)$.%, for some computable function $g$ on $k$ and $\epsilon$. 
\end{theorem}

\subsection{Our Techniques} 
\paragraph{Overview of FPT Time $(3+\epsilon)$-Approximation for Minimum-Norm Capacitated $k$-Clustering} In the overview, we mainly focus on the $\topell$ norm capacitated $k$-clustering problem, as it already captures the essence of our algorithm. The problem generalizes both capacitated $k$-center (with $\ell = 1$) and capacitated $k$-median (with $\ell = n$). Letting $\ell$ go from $1$ to $n$ gives a smooth transition from the $k$-center to the $k$-median objective. Therefore, we need to achieve $(3+\epsilon)$-approximation algorithms for both extreme cases while unifying their key ideas to handle the general $\ell$ case.

Our FPT time $3$-approximation for capacitated $k$-center is new, and we sketch the intuition. For simplicity, we first focus on the soft-capacitated setting, where facilities may be opened multiple times. (The total number of open facilities, counting multiplicities, remains $k$.) 

By guessing the optimum value, scaling and rounding distances, we can assume all distances are integers and the optimum value is $1$. We construct a solution $S$ with cost $1$ and $O(k \log n)$ open facilities that respect the capacity constraints; this can be easiliy obtained via LP rounding. For each facility in $S$, we sample $O\left(\log n\right)$ clients connected to it.

For an optimum cluster $(i^*, J)$,  $i^*$ is the open facility and $J$ is the assigned clients. If every $j \in J$ is connected to a facility $i$ with $u_i \leq |J|$ in the solution $S$, then with high probability, some client in $J$ will be sampled. This sampled client $j$ can then serve as a \emph{pivot} for the cluster: we open the facility with the largest capacity within distance at most $1$ of $j$. 

On the other hand, if some $j \in J$ is connected to a facility $i \in S$ with capacity $u_i > |J|$, then we can directly use $i$ as a replacement for $i^*$. In either case, by guessing, we can identify a facility $i\in S$ with $u_i \geq |J|$ and $d(i, i^*) \leq 2$, which leads to a 3-approximation.

The hard-capacitated case can be addressed using the coloring idea~\cite{cohen2022fixed}, along with additional care. By guessing a coloring function, we can ensure with reasonable probability that the optimum facilities have distinct colors, which can be used to guide our selection of facilities. We have a stricter condition for the second case above: We require that $i \in S$ has capacity $u_i \geq k|J|$ so that $i$ have enough capacity to serve $k$ clusters in the optimum solution.  We also need to take care of the case where an optimum facility also appears in $S$.

Our algorithm for the $\topell$ norm needs to capture an FPT time $(3+\epsilon)$-approximation for capacitated $k$-median. Such an algorithm was known \cite{cohen2022fixed}. It relies on the coreset technique, which has been extensively studied in the literature \cite{conf/icml/BravermanJKW19, conf/focs/BravermanCJKST022, conf/soda/Huang0L025, Chen2006, Chen2009, Feldman2020}. However, the technique is based on sampling and only works well when $\ell = \Omega(n)$. This makes the algorithm hard to combine with the $k$-center objective.   Instead, we use an alternative $(3+\epsilon)$-approximation for capacitated $k$-median that avoids coresets. 

We obtain a $(1+\epsilon)$-approximation for the problem using $O(\frac{k\log n}\epsilon)$ facilities $S$. We sample a few clients from each cluster in solution $S$. Additionally, we sample a few clients in $C$, with probabilities proportional to their costs in $S$. If for an optimum cluster $(i^*, J)$, the total cost of $J$ in the solution $S$ is large, we likely sampled a ``good'' client in $J$, which can serve as a pivot. Otherwise, the total cost of $J$ in the solution $S$ can be essentially ignored, and some facility with large enough capacity in $S$ can be used to replace $i$. Again, extra-care is needed to handle hard capacities.

To unify the approaches for both the capacitated $k$-center and $k$-median problems, we guess the $\ell$-th largest connection distance $t$ in the optimum solution. The $\topell$ norm of the optimum distance vector $v \in \R_{\geq 0}^C$ is $\ell t + \sum_{j \in C}(v_j-t)^+$. At a very high level, we treat the $\ell t$ term as the $k$-center component, and the $\sum_{j \in C}(v_j-t)^+$ term as the $k$-median component. Our algorithm combines the techniques to handle these two parts effectively.

To extend the idea to a general monotone symmetric norm $f$, we apply the above idea for all distances $t$ that is an integer power of $1+\epsilon$.  This will bound the top-$\ell$ norm cost of the solution, for every $\ell \in [0, n]$. Every symmetric norm is the maximum of many \emph{order norms}, which in turn is a non-negative linear combination of top-$\ell$ norms \cite{Chakrabarty2018}. Therefore, a bound on the top-$\ell$ norms leads to a bound on the $f$-norm. This proves Theorem~\ref{thm:topell-kC}. This technique has been used in the literature for many other problems \cite{DBLP:journals/dam/Tamir01, DBLP:journals/mp/AouadS19}.

\paragraph{Overview of $\Big(1 + \frac{2}{ce} + \epsilon\Big)$-Approximation for Top-$cn$ Norm (Uncapacitated) $k$-Clustering Problem for $c \in (\frac1 e, 1]$} To describe our algorithm, we first give an overview of the $(1+\frac2e)$-approximation for $k$-median in FPT time by \cite{DBLP:Cohenaddad2019}. By using the coreset technique, one can assume that the set of clients is located in only $O\!\big(\frac{k \log n}{\epsilon}\big)$ positions. Given a cluster $(i^*, J)$ in the optimum clustering $\mathcal{C}$, a \emph{leader} of the cluster is defined as the client in $J$ that is nearest to $i^*$.

The algorithm proceeds by guessing the $k$ leaders and the approximate distances of the leaders to their respective centers. With this information, a $(1 + \frac2e)$-approximation can be obtained via submodular maximization. Roughly speaking, we consider a ball centered at each leader, with the radius equal to the guessed distance to its center. Ensuring that every ball contains one facility yields a cost of $3 \cdot \opt$. Choosing the optimum facilities within these balls improves the cost by $2 \cdot \opt$. The goal then becomes maximizing this improvement, which can be cast as a submodular maximization problem under a partition matroid constraint. The standard $(1 - \frac1e)$-approximation ratio for submodular maximization translates into an approximation ratio of $3 - 2(1 - \frac1e) = 1 + \frac2e$ for $k$-median.

%{\color{red}Algorithms based on linear programming and rounding strategies have been widely investigated for various clustering problems \cite{DBLP:conf/icalp/CharikarL12, DBLP:conf/focs/PalTW01, DBLP:journals/mp/ChudakW05}.}

To generalize this approach to the $\topcn$ setting, we modify the algorithm in two ways. First, one would need to construct a coreset for the $\topcn$ objective. We believe this is possible for any constant $c \in (0, 1)$; however, we instead adopt an alternative approach that is similar to our algorithm in the capacitated setting and avoids the use of coresets. We obtain a bi-criteria $(1 + \epsilon)$-approximation for the problem by opening $O\big(\frac{k \log n}{\epsilon}\big)$ facilities $S$. We sample a few clients $R$ with probability proportional to their costs in $S$, and then guess a set of $k$ pivots from $R \cup S$. Pivots play a role similar to leaders but can be any points in the metric space, not necessarily the clients in the respective optimum clusters. With reasonable probability, when the guesses are correct, the pivots have the desired property that leaders have.  

The second modification is essential. With the $\topcn$ cost, the improvement function is no longer submodular. Instead, we use an LP-based approach to construct the set of centers. A useful notion here is the \emph{occurrence time vector} of a cost vector: in such a vector $\delta$, $\delta_t$ represents the number of times distance $t$ occurs in the cost vector. We naturally allow occurrence times to be fractional. We then formulate an LP relaxation that uses the pivot and radius information, with the objective of minimizing the $\topcn$ cost of the occurrence time vector $\delta$.

We design a randomized rounding algorithm such that, in expectation, the occurrence time vector of the integral solution is no worse than $(1 - \frac1e)\delta + \frac{1}{e}(\delta \otimes 3)$, where $\delta \otimes 3$ denotes the vector obtained from $\delta$ by scaling all distances by a factor of 3. We bound the ratio of the $\topcn$ cost of the vector to that of $\delta$. The worst-case scenario arises when the cost vector is the all-one vector, i.e., when the occurrence time vector has distance 1 appearing $n$ times and all other distances appearing zero times. This yields a bound of $1 + \frac{2}{ec}$. Finally, using the occurrence time representation, the $\topcn$ function is concave, leading to an overall approximation ratio of $1 + \frac{2}{ec} + \epsilon$. This proves Theorem~\ref{theo: top-cn norm}.

A simple modification to the algorithm can handle the $L_\infty$ norm, and simultaneously provide a $3$-approximation for it, proving Theorem~\ref{thm:bi-criteria}.

\subsection{Other Related Work}
For the $k$-supplier-center problem, the special case where $F = C$ has been studied more in the literature and is simply known as the $k$-center problem. Simple greedy algorithms achieve $2$ and $3$-approximations for the $k$-center and $k$-supplier-center problems respectively~\cite{hochbaum1985best, Plesnik1987}. They are the best possible for the two problems assuming $\text{P} \neq \text{NP}$.

Charikar, Guha,  Tardos, and  Shmoys \cite{Charikar2002} gave the first constant factor approximation algorithms for the $k$-median problem, which is a $6\frac23$-approximation based on LP rounding. Since then, there has been a long line of work aiming at improving the approximation ratio \cite{arya2001local, byrka2017improved, Charikar1999, cohen2022improved}. 
In a recent milestone result, Cohen-Addad, Grandoni, Lee, Schwiegelshohn and Svensson \cite{cohen20252+}, using the Lagrangian Multiplier Preserving (LMP) $2$-approximation for the uncapacitated facility location problem due to Jain and Vazirani \cite{jain2001approximation}, together with a novel framework for converting this algorithm into one for $k$-median with only a loss of $1+\epsilon$ in the approximation ratio.

For the metric $k$-means problem, Gupta and Tangwongsan~\cite{gupta2008simpler} provided the first $O(1)$-approximation, which is a $16$-approximation based on local search. This factor was later improved by \cite{Ahmadian2020} to $9 + \epsilon$.  Recently,  Charikar,  Cohen-Addad, Gao,  Grandoni, Lee and Wijland \cite{DBLP:CharikarCGGLW25} gave  a $3+\sqrt{2}+\epsilon$-approximation algorithm, improving previous work.
A more commonly used variant in unsupervised machine learning, which is simply referred to as the $k$-means problem, considers the case where $C \subseteq F = \mathbb{R}^\ell$ and $d$ is the Euclidean metric. A widely used heuristic for the problem is Lloyd's algorithm \cite{lloyd1982least}.  
% Kanungo  Mount, Netanyahu, Piatko, 
% Silverman and Wu~
For $k$-means, \cite{kanungo2002} gave a $(9+\epsilon)$-approximation algorithm, which was improved to $6.357$ by \cite{Ahmadian2020}.

For the capacitated $k$-median problem, there is a line of work that provides polynomial time bi-criteria approximation algorithms, which either violate the number $k$ of open facilities \cite{aardal2015approximation}, or the capacity constraints \cite{chuzhoy2005approximating, li2014improved, byrka2014bi}. These algorithms are based on the natural LP relaxation for the problem. There are gap instances showing that, in order to obtain an $O(1)$-approximation, the violation factors for the number of open facilities and the capacities constraints must be at least $2$.  This barrier was overcome by Li \cite{Li2017uniform}, who gave an $O(1)$-approximation for the problem with uniform capacities by opening $(1+\epsilon)k$ facilities, using a stronger LP relaxation. This result has since been generalized to the non-uniform setting \cite{Li2016}. For $O(1)$-approximations with capacity violations, the violation factor has also been reduced to $(1+\epsilon)$ in both the uniform-capacitated \cite{byrka2016approximation} and non-uniform capacitated \cite{demirci2016constant} versions of the problem. 

Adamczyk, Byrka, Marcinkowski, Meesum and Wlodarczyk \cite{Adamczyk2019} gave a $(7+\epsilon)$-approximation for the capacitated $k$-median problem in FPT time, which is the first $O(1)$-approximation for the problem in the FPT setting. This approximation factor was later improved to $3+\epsilon$ by \cite{cohen2022fixed}, who also provided a $(1+\epsilon)$-approximation for the problem in  Euclidean spaces of arbitrary dimensions. Both algorithms run in FPT time. \cite{xu2019constant} presented an FPT-time $(69+\epsilon)$-approximation algorithm for the hard-capacitated $k$-means problem in Euclidean spaces. Some other variants of $k$-clustering problems in the FPT-time region were also studied \cite{Goyal2023, Feng2020, Bandyapadhyay2023}. Recently, Liu,  Chen and Xu \cite{liu2025fixed} studied the uniform capacitated $k$-supplier problem in FPT time and obtained a 5-approximation algorithm.

\paragraph{Organization} In Section~\ref{sec:prelim}, we give some preliminaries.  In Section~\ref{sec:top-t-MNCkC}, we give the $(3+\epsilon)$-approximation for the minimum-norm capacitated $k$-clustering problem in FPT time, proving Theorem~\ref{thm:topell-kC}. %We generalize the algorithm to an arbitrary monotone norm $f$ in Section~\ref{sec:general-norm-CkC}, proving Theorem~\ref{thm:general-norm-CkC}. %$(3+\epsilon)$-approximation for minimum-norm capacitated $k$-clustering in FPT time similarly.
In Section~\ref{sec:top-cn-kC}, we give the tight $\big(\min\big\{1 + \frac2{ec}, 3\big\} + \epsilon\big)$-approximation algorithm for the $\topcn$ norm $k$-clustering problem, proving Theorem~\ref{theo: top-cn norm}. 
All missing proofs can be found in the appendix; in particular,  we defer the proof of Theorem~\ref{thm:bi-criteria} to Appendix~\ref{sec:2-norms}.

    \section{Preliminaries}
\label{sec:prelim}

For any $a \in \R$, we let $(a)^+ = \max\{a, 0\}$. For every real-valued vector $y$ over some domain $U$ and any $S\subseteq U$, we let $y(S):= \sum_{i \in S}y_i$, unless otherwise defined.  

Given a metric space $(V,d)$,  $v\in V$ and $S\subseteq V$,  we define $d(v, S):=\min_{i\in S}d(v, i)$ to be the distance from $v$ to the set $S$. For any subset $U \subseteq V, v \in V$ and $r \geq 0$, we define $\ball_U(v, r):= \{u \in U: d(u, v) \leq r\}$ to be the set of points  in $U$ with distance at most $r$ to $v$. 

\subsection{Norms}
       \begin{definition}[Norms]  
       %\shi{Better to separate the definitions of norms, monotone and symmetric norms.}
        \label{def:norm}
       A function $f: \R_{\geq 0}^n\to \R_{\geq 0}$ is a norm, if it satisfies the following 3 properties:
       \begin{enumerate}[(\ref{def:norm}a)]
           \item $f({x})=0$ iff ${x}=0$ (non-negativity).
        \item For any ${x}\in \R_{\geq 0}^n, \lambda\in \R_{\geq 0}$, we have
           $f(\lambda {x})=\lambda f({x})$ (homogeneity).
        \item For any ${x}, {y}\in \R_{\geq 0}^n$, we have $f({x}+{y})\leq f({x})+f({y})$ (subadditivity).
       \end{enumerate}
\end{definition}

For a norm $f$, we shall also use $|x|_f$ to denote $f(x)$, the $f$-norm of the vector $x$. 

\begin{definition}[Monotone and Symmetric Norms]
%\shi{Maybe both words are correct. I always used ``monotone".}
   A norm $f:\R_{\geq 0}^n \to \R_{\geq 0}$ is said to be monotone, if for any ${x}, {y}\in \R_{\geq 0}^n$ satisfying ${x}\leq {y}$, we have $f({x})\leq f({y})$. 
    It is said to be symmetric, if for any $x \in \R_{\geq 0}^n$, and any permutation matrix $A \in \{0, 1\}^{n \times n}$, we have $f(x) = f(Ax)$. 
   %i.e., for any ${x}\in \R^n$, any permutation of the coordinates of ${x}$ does not change the value of the function, namely $f({x})=f(\boldsymbol{x'})$, where $\boldsymbol{x'}$ is the vector obtained by arbitrarily permuting the coordinates of ${x}$.   
\end{definition}

Here are some common monotone symmetric norms studied in the literature:
\begin{itemize}
    \item $L_p$ norms, $p\geq 1$: $L_p(x) := \left(\sum_{j = 1}^n x_j^p\right)^{1/p}, \forall x \in \R_{\geq 0}^n$. We also use $|x|_p$ for $L_p(x)$.
    \item $L_\infty$ norm: $L_\infty(x) := \max_{j \in [n]} x_j, \forall x \in \R_{\geq 0}^n$. We also use $|x|_\infty$ for $L_\infty(x)$.
    \item Top-$\ell$ norm, $\ell \in [0, n]$ is a real: $|x|_\topell$ for any $x\in \R_{\geq 0}^n$ is the maximum of $\alpha^\rmT x$ over all $\alpha \in [0, 1]^n$ satisfying $|\alpha|_1 = \ell$. When $\ell$ is an integer, $x_\topell$ is the sum of the $\ell$ largest coordinates in $x$. In this paper, we shall allow $\ell$ to take fractional values. Notice that $\topnorm{1} \equiv L_\infty$ and  $\topnorm{n} \equiv L_1$.
    \item Ordered norms: This is a non-negative linear combination of top-$\ell$ norms, for different values of $\ell$. See the following definition.
\end{itemize}

% Every ordered norm is a nonnegative linear combination of Top-$\ell$ norms.
\begin{definition}[Ordered norms]
    Given a non-negative and non-increasing weight vector $w = (w_1, \dots, w_n)$. The $w$-ordered norm of any vector $x \in \R_{\geq 0}^n$, denoted as $|x|_{w\text{-}\mathtt{ordered}}$, is defined as follows:
    \begin{align*}
        |x|_{w\text{-}\mathtt{ordered}}:= \sum_{i = 1}^n w_i x^\downarrow_i,
    \end{align*}
    where $x^\downarrow_i$ is the $i$-th largest value in $\{x_1, x_2, \cdots, x_n\}$ (counting multiplicities). 
\end{definition}

The following folklore lemma is useful when analyzing top-$\ell$ norms:
\begin{lemma}
    \label{lemma:linearize-top-l}
    For any $x \in \R_{\geq 0}^n$, we have $|x|_\topell = \min_{t \geq 0}\left(\sum_{j = 1}^n (x_j-t)^+ +  \ell t \right)$. Moreover, the minimum is achieved when $t$ is the $\ell$-th largest coordinate in $x$. 
\end{lemma}

The following observation is easy to see:
\begin{observation}
    Given a non-negative and non-increasing weight vector $w = (w_1, \dots, w_n)$. Let $w_{n+1}=0$. Then for any $x \in \R_{\geq 0}^n$, we have $|x|_{w\text{-}\mathtt{ordered}} =\sum_{\ell=1}^n(w_\ell-w_{\ell+1})\cdot |x|_\topell$.
\end{observation}

Any monotone symmetric norm can be expressed as the maximum of ordered norms \cite{Chakrabarty2018}:
    \begin{lemma}[\cite{Chakrabarty2018}]
        \label{lemma:symmetric-to-top}
        Let $f:\R_{\geq 0}^n \to \R_{\geq 0}$ be a monotone symmetric norm. There exists a closed set $\mathcal{W}$ of non-increasing vectors in $\R_{\geq 0}^n$, such that for any $x\in \R^n_{\geq 0}$, we have \begin{align*}
            |x|_f =\max_{w\in \mathcal{W}}|x|_{w\text{-}\mathtt{ordered}} =\max_{w\in \mathcal{W}}\sum_{\ell=1}^n(w_\ell - w_{\ell+1})|x|_{\topell}.
        \end{align*}
    \end{lemma}

\subsection{Definitions of Problems} 
    \begin{definition}[Minimum-Norm $k$-Clustering] \label{def: $k$-Clustering under $f$-Norm}
        Let $f:\R_{\geq 0}^n \to \R_{\geq 0}$ be a monotone symmetric norm.  In the Minimum-Norm $k$-Clustering problem under $f$-norm, we are given a set $F$ of facilities,  a set $C$ of $n$ clients, a positive integer $k \leq |F|$, and a metric space $(F \uplus C, d)$ over $F \uplus C$. The goal of the problem is to find a set $S \subseteq F$ with $|S| = k$ so as to minimize $f\big((d(j, S))_{j \in C}\big)$. 
    \end{definition}

    When $f$ are the $L_\infty, L_1$ and $L_2$ norms, the problem becomes $k$-supplier-center, $k$-median and metric $k$-means problems respectively.

    \begin{definition}[Minimum-Norm Capacitated $k$-Clustering] \label{def:MNCk}
        Let $f:\R_{\geq 0}^n \to \R_{\geq 0}$ be a monotone symmetric norm.  In the Minimum-Norm Capacitated $k$-Clustering problem under $f$-norm, we are given $F, C, n, k$ and $d$ as in Definition~\ref{def: $k$-Clustering under $f$-Norm}. Additionally, we are given a capacity $u_i \in \Z_{\geq 0}$ for every $i \in F$. The goal of the problem is to find a set $S \subseteq F$ with $|S| = k$, and an assignment vector $\sigma \in S^C$ such that $|\sigma^{-1}(i)|\leq u_i$ for every $i\in S$, so as to minimize $f\big((d(j, \sigma_j))_{j \in C}\big)$. 
    \end{definition}

    When $f$ are the $L_\infty$ and $L_1$ norms, the problem becomes capacitated $k$-center\footnote{In the literature, capacitated $k$-center refers to the problem where $F$ and $C$ can be unrelated, instead of the problem where $F = C$.} and capacitated $k$-median respectively.  \medskip

    Throughout the paper, we assume we are given an efficient oracle for the norm $f$. We can assume distances are either $\infty$ or integers in $[0, \poly(n)]$, by losing a factor of $1+\epsilon$ in the approximation ratio. See Appendix~\ref{appendix:distance-polynomial}. As is typical, we do sampling and make guesses during our algorithms. When analyzing the algorithm, we define the conditions for the sampling steps being successful, and the correct answers for all the guesses.  In the actual algorithm, we have to enumerate all possibilities for the guesses.  It suffices for our algorithm to succeed with $\frac1{g(k, \epsilon)\poly(n)}$ probability, as we can repeat it many times and output the best solution generated, to increase the probability to $1-1/\poly(n)$. For the guessing and repetition to be feasible, it is required that we can compute the cost of the constructed solutions. For most natural norms, the optimal assignment $\sigma$, and thus the minimum cost, can be computed efficiently given the set $S$ of open facilities. For general norms $f$, the cost can be approximated within a factor of $1+\epsilon$.

\subsection{$(1+\epsilon)$-Approximation for Capacitated $k$-Clustering with $O(\frac{k\log n}{\epsilon})$ Facilities}
It is useful to run an LP rounding algorithm to obtain a $(1+\epsilon)$-approximate solution for capacitated $k$-clustering with $O(\frac{k\log n}{\epsilon})$ facilities $S$. With FPT time allowed, we can afford to guess $\poly(k, \frac1\epsilon)$ facilities in $S$. 

We are given $F, C, n, d, k$ and capacities $(u_i)_{i \in F}$ as in a capacitated $k$-clustering problem. Instead of a monotone symmetric norm $f$, we are given a convex monotone function $h: \R_{\geq 0}^C \to \R_{\geq 0}$ (which is not necessarily symmetric or a norm). The goal is the same as that of the minimum-norm capacitated $k$-clustering problem, except that we are minimizing the $h$ function of the connection distance vector. %Consider the problem of finding a set $S \subseteq F$ with $|S| = k$ and an assignment vector $\sigma \in S^C$ with $|\sigma^{-1}(i)| \leq u_i, \forall i \in S$, so as to minimize $h\big((d(j, \sigma_j)_{j \in C}\big)$. 
Let $\opt'$ be the value of the instance.

For some $i \in F$ and $J \subseteq C$, we say $(i, J)$ is valid star if $|J| \leq u_i$. The following theorem summarizes the $(1+\epsilon)$-approximation, whose proof is deferred to Appendix~\ref{appendix:pseudo-approx}. 
\begin{restatable}{theorem}{MainThm}  \label{thm:pseudo-approx}
    Let $\epsilon > 0$ be a constant. We can efficiently find a set $\calS$ of $O\left(\frac{k \ln n}{\epsilon}\right)$ valid stars such that
    \begin{itemize}
        \item $\biguplus_{(i, J) \in \calS} J = C$.
        \item  Let $b\in \mathbb R_{\geq 0}^C$ be the connection distance vector induced by $\calS$:
        For every $j \in C$, we have $b_j = d(i, j)$ for the unique star $(i, J)$ with $j \in J$. Then $h\left((1-\epsilon)b\right) \leq \opt'$.
    \end{itemize}
    The algorithm succeeds with high probability. %When $f$ is the $\ell_\infty$ norm, the size of $\calS$ can be bounded by $O\left(k \log n\right)$ and we can guarantee $f(b) \leq \opt$ in the second property.
\end{restatable}

The following corollary is an immediate consequence of our analysis, and we were unable to find an explicit statement of the result in the literature. As we discussed, it was implied by \cite{tcs/GoyalJ23} when all facilities have uniform capacities. However, extending their result to the general capacitated setting seems non-trivial. For the $\topcn$ norm $k$-clustering problem, we need to use the corollary to handle the case $c \leq \frac{1}{e}$. We defer the proof to Appendix~\ref{appendix:pseudo-approx}.
\begin{restatable}{corollary}{corothreeapprox}
    \label{coro:three-approximation}
    Consider the minimum-norm $k$-clustering problem under a monotone norm $f$ (which is not necessarily symmetric). 
    There is a $(3+\epsilon)$-approximation algorithm for the problem in FPT time with parameters $k$ and $\epsilon$. 
\end{restatable}

\subsection{Finding $(1+\epsilon)$-Approximate Assignment for Minimum-Norm Capacitated $k$-Clustering with Open Facilities}

Consider an instance of the Minimum-Norm Capacitated $k$-Clustering problem.  For many natural norms such as $L_p$ and $\topnorm{\ell}$ norms, when given the set $S$ of $k$ open facilities, the optimum assignment of $C$ to $S$ can be found easily. However, this is not trivial for general monontone symmetric norms $f$. Instead, we present an FPT-time algorithm to find a $(1+\epsilon)$-approximate assignment. The proof of the following theorem is described in Appendix~\ref{appendix:find-assignment}. 

\begin{restatable}{theorem}{thmfindassignment}
    \label{thm:find-assignment}
    Consider a minimum-norm capacitated $k$-clustering instance defined by $F, C, n, d, u$ and $k$, under a symmetric monotone norm $f$. 
     Assume $|F|  = k$
     and $\epsilon > 0$ is a constant. We can find a $(1+\epsilon)$-approximate assignment $\sigma \in F^C$ for the instance in time $g(k, \epsilon) \cdot \poly(n)$, for a computable function $g$ depending on $k$ and $\epsilon$.
\end{restatable}

    \newcommand{\capacity}{\mathrm{cap}}
\section{FPT Time $(3+\epsilon)$-Approximation for Minimum-Norm Capacitated $k$-Clustering}
\label{sec:top-t-MNCkC}

In this section, we give the FPT time $(3+\epsilon)$-approximation algorithm for the minimum-norm capacitated  $k$-clustering problem, proving Theorem~\ref{thm:topell-kC}. Recall that we say $(i \in F, J \subseteq C)$ is a valid star if $|J| \leq u_i$. We fix an optimum solution $\calS^*$ of $k$ valid stars that is unknown to the algorithm. $\calS^*$ cover all clients in $C$, and all facilities are distinct. Let $S^*$ be the set of facilities in $\calS^*$. So, we have $|S^*|=|\calS^*| = k$. Let $\opt$ be the $f$-norm cost of $\calS^*$.

\begin{algorithm}[h]
    \caption{$(3+\epsilon)$-Approximation for Minimum-Norm Capacitated $k$-Clustering}
    \label{algo:MNCkC}
    \begin{algorithmic}[1]
        \State define set $\bfT$ of non-negative reals with $|\bfT| \leq O\left(\frac{\log n}{\epsilon}\right)$, as described in text \label{step:MNCkC-T}
        \label{step:MNCkC-bfT}
        \For{every $t \in \bfT$} %\Comment{for each $t$, construct a top-$t$ norm capacitated $k$-clustering instance}
            \State with objective $h(v) := \sum_{j \in C}(v_j - t)^+$ for every $v \in \R_{\geq 0}^C$, use Theorem~\ref{thm:pseudo-approx} to obtain a set $\calS^t$ of $O\left(\frac{k \ln n}{\epsilon}\right)$ valid stars \label{step:MNCkC-calS}
            \State $S^t \gets \{i \in F: \exists J, (i, J) \in \calS^t\}$ \label{step:MNCkC-S}
        \EndFor
        %\State $S \leftarrow \cup_{t \in \bfT} S^t, R \leftarrow \cup_{t \in \bfT} R^t$ 
        \State randomly choose a color function $\mathtt{color}: F \to [k]$ \label{step:MNCkC-color}
        \State guess the types of each color $c \in [k]$ \Comment{each color is of type-1, 2 or 3} \label{step:MNCkC-type}
        \For{every $t \in \bfT$}
         $R^t \leftarrow \texttt{MNCkC-choose-R}(t)$ \label{step:MNCkC-R} \Comment{clients in $R^t$ are called \emph{representatives}} 
        \EndFor
        \State $S \leftarrow \union_{t \in \bfT} S^t, R \leftarrow \union_{t \in \bfT} R^t$ \label{step:MNCkC-merge}
        \State guess a pivot $p_c$ for each type-1 or 2 color $c \in [k]$, and $i^*_c$ for each type-3 color $c \in [k]$ such that \label{step:MNCkC-pivots} 
        \Statex \Comment{$p_c$'s for type-1 and 2 colors $c$, $i^*_c$'s for all colors $c \in [k]$ are defined in text}
        \begin{itemize}
            \item $p_c \in R$ if $c$ is of type-1, $p_c \in S$ if $c$ is of type-2, and
            \item $i^*_c \in S$ has color $c$ for a type-3 color $c$.
        \end{itemize}
        \State guess a $(1+\epsilon)$-approximate overestimation $r_c$ for $d(i^*_c, p_c)$ for every type-1 or 2 color $c$ \label{step:MNCkC-radius}
        \State \Return $\texttt{MNCkC-clustering-with-pivots}()$ \Comment{See Algorithm \ref{algo:MNCkC-clustering-with-pivots} for its definition} \label{step:MNCkC-return}
    \end{algorithmic}
\end{algorithm}

The algorithm is stated in Algorithm \ref{algo:MNCkC}. In Step~\ref{step:MNCkC-T}, we define a set $\bfT$ of distances depending on the problem, as follows:
\begin{itemize}
    \item If the norm to minimize is the $\topnorm{\ell}$ norm, then let $\bfT = \{t\}$, with $t$ being the $\ell$-th largest connection distance in the optimum solution $\calS^*$. There are $n|F|$ possibilities for $t$, so it is affordable to guess its correct value. 
    \item If the norm is a general monotone symmetric norm, then $\bfT =\Big\{(1+\epsilon)^{\ceil{\log_{1+\epsilon} d(i,j)}} : i \in F, j \in C\Big\}$. Notice that $|\bfT| = O(\frac{\log n}{\epsilon})$ as we assumed that distances that are not $\infty$ are integers bounded by $\poly(n)$. 
\end{itemize}
Readers seeking for a more efficient understanding of the core ideas can focus on the $\topell$ norm case, where we have only one $t$ in $\bfT$. 

For every $t \in \bfT$, in Step~\ref{step:MNCkC-calS}, we obtain a set $\calS^t$ of $O\left(\frac{k \ln n}\epsilon\right)$ valid stars covering $C$ using Theorem~\ref{thm:pseudo-approx}, and define $S^t$ to be the set of facilities used in $\calS^t$ in Step~\ref{step:MNCkC-S}. We let $i^t_j$ be the center of the star in $\calS^t$ containing $j$ for every $j \in C$, and let $b^t_j = d(j, i^t_j)$ be the connection distance of $j$ in $\calS^t$. The conditions of the Theorem~\ref{thm:pseudo-approx} hold with high probability. We assume they are satisfied: 

 \begin{itemize}
      \item[(P)] For every $t \in \bfT$, we have $\sum_{j \in C}((1-\epsilon)b^t_j - t)^+ \leq \sum_{(i^*, J) \in \calS^*, j \in J}(d(i^*, j) - t)^+$.
 \end{itemize}

%For every $(i, J) \in \calS$, we have $|J| \leq u_i$ and $d(i, j) \leq \opt, \forall j \in J$. %\textcolor{red}{(Sijin: A minor problem - Thm \ref{thm:pseudo-approx} only proves for $\ell_\infty$, rather than top-$\ell$ norm.)} 
%Every $j$ appears in exactly one star in $\calS$. 

% In Step~\ref{step:MNCkC-merge}, the algorithm takes the union of all $S^t$ for $t \in \bfT$ to get $S$ for this instance. We have $|S| = O(\frac{k \ln^2 n}{\epsilon^2})$.

We describe the remaining steps in more detail.

\subsection{Steps~\ref{step:MNCkC-color} and \ref{step:MNCkC-type} of Algorithm~\ref{algo:MNCkC}: Guessing Colors and Types}
In Step~\ref{step:MNCkC-color} of Algorithm~\ref{algo:MNCkC}, we randomly choose a function $\mathtt{color}: F \to [k]$. With probability $\frac{k!}{k^k}$, the $k$ facilities in $S^*$ have distinct colors. We assume this happens. For every color $c \in [k]$, let $(i^*_c, J^*_c)$ be the star in $\calS^*$ such that $i^*_c$ is of color $c$. We define $\bar J^*_c$ to be the $\ceil{\epsilon |J^*_c|}$ clients in $J^*_c$ closest to $i^*_c$. Let $\tilde J^*_{t,c}$ be the $\ceil{\frac{|\bar J^*_c|}{2}}$ clients in $\bar J^*_c$ with smallest $b^t_j$ values for every $t \in \bfT$. Notice that the $i^*_c$'s, $J^*_c$'s, $\bar J^*_c$'s and $\tilde J^*_{t, c}$'s are not known to our algorithm after the steps.

%TODO: change the objective of the LP in prelminary

We define a \emph{type} for each color $c \in [k]$ as follows:
% \remove{\begin{itemize}
%     \item If $i^*_c \notin S$, then
%     \begin{itemize}
%         \item if $\forall j \in \tilde J^*_{t,c}, u_{i^t_j} < k|J^*_c|$, then $c$ is of \emph{type-1a}, otherwise,
%         \item if $\sum_{j \in \bar J^*_c}((1-\epsilon)b^t_j - t)^+ \geq \frac{\epsilon^2}{k}\cdot \sum_{j \in C}((1-\epsilon)b^t_j-t)^+$, then $c$ is of \emph{type-1b}, and otherwise,
%         \item $c$ is of \emph{type-2}.
%     \end{itemize}
%     \item If $i^*_c \in S$, then $c$ is of \emph{type-3}.
% \end{itemize}}
\begin{itemize}
    \item If $i^*_c \notin S$, then
    \begin{itemize}
        \item if $\exists t \in \bfT, \forall j \in \tilde J^*_{t,c}, u_{i^t_j} < k|J^*_c|$, then $c$ is of \emph{type-1a}, otherwise,
        \item if $\exists t \in \bfT,\sum_{j \in \bar J^*_c}((1-\epsilon)b^t_j - t)^+ \geq \frac{\epsilon^2}{k}\cdot \sum_{j \in C}((1-\epsilon)b^t_j-t)^+$, then $c$ is of \emph{type-1b}, and otherwise,
        \item $c$ is of \emph{type-2}.
    \end{itemize}
    \item If $i^*_c \in S$, then $c$ is of \emph{type-3}.
\end{itemize}

Notice that for the type-1a colors, we consider the set $\tilde J^*_{t, c}$, but for type-1b colors, we consider the set $\bar J^*_c$.  We say $c$ is of \emph{type-1} if it is of type-1a or type-1b. Notice the types 1, 2 and 3 partition the set $[k]$ of colors.  We guess the types in Step~\ref{step:MNCkC-type} of Algorithm~\ref{algo:MNCkC}. Again, we assume our guesses are correct.

\subsection{Step~\ref{step:MNCkC-R} of Algorithm~\ref{algo:MNCkC}: Constructing Representatives $R^t$}

In Step~\ref{step:MNCkC-R} of Algorithm~\ref{algo:MNCkC}, for every $t \in \bfT$, we construct a \emph{representative set} $R^t$ through the process $\texttt{MNCkC-choose-R}(t)$, described in Algorithm~\ref{algo:top-t-cap-choose-R}.

\begin{algorithm}[h]
    \caption{$\texttt{MNCkC-choose-R}(t)$}
    \label{algo:top-t-cap-choose-R}
    \begin{algorithmic}[1]
        \State for every $(i, J) \in \calS^t$, uniformly sample $\min(|J|,\ceil{\frac{2k}{\epsilon} \ln (kn)})$ clients from $J$, without replacement; they constitute set $R^t_1$
        \State sample $O\left(\frac{k\ln n}{\epsilon^2}\right)$ clients from $C$, for a sufficiently large hidden constant, each client $j$ being chosen from $C$ with probabilities proportional to $((1-\epsilon)b^t_j-t)^+$, with replacement; they constitute set $R_2^t$
        \State \Return $R^t = R^t_1 \cup R_2^t$
    \end{algorithmic}
\end{algorithm}

As $|\calS^t| = O\left(\frac{k \ln n}{\epsilon}\right)$, we have $|R^t| = O\left(\frac{k^2\ln^2 n}{\epsilon^2}\right)$ for every $t \in \bfT$.  

In Step~\ref{step:MNCkC-merge}, we merge all sets $S^t$ into $S$ and $R^t$ into $R$. Therefore, $|S| = O\left(\frac{k \log^2 n}{\epsilon^2}\right)$ and $|R| = O\left(\frac{k^2\log^3 n}{\epsilon^3}\right)$.

\begin{lemma}\label{lem:existence-ra}
    With probability at least $1-\frac{1}{n}$, the following event happens: for every color $c \in [k]$ of type-1, we have $\bar J^*_c \cap R \neq \emptyset$.
\end{lemma}
\begin{proof}
    Fix a type-1a color $c$. Focus on the $t \in \bfT$ that makes it of type-1a. So, $\forall j \in \tilde J^*_{t, c}$, we have $u_{i^t_j} < k|J^*_c|$. As $(i, J) \in \calS$ satisfies $|J| \leq u_i$, we have $\Pr[j \in R^t_1] \geq \min\Big\{1, \frac{2k\ln (kn)}{\epsilon\cdot u_{i^t_j}}\Big\}$ for every $j \in \tilde J^*_{t, c}$.
    
   If for some $j \in \tilde J^*_{t,c}$ we have $\Pr[j \in R^t_1] = 1$, then $\Pr[\tilde J^*_{t,c} \cap R^t_1 \neq \emptyset] = 1$. So, we assume for every $j \in \tilde J^*_{t,c}$, we have $\Pr[j \in R^t_1] \geq \frac{2k\ln (kn)}{\epsilon \cdot u_{i^t_j}}\geq \frac{2\ln (kn)}{\epsilon \cdot |J^*_c|} \geq \frac{\ln (kn)}{|\tilde J_{t, c}^*|}$. The first inequality holds as $u_{i^t_j} < k|J^*_c|$, and the second inequality follows from that $|\tilde J^*_{t, c}| \geq \frac{1}{2}\cdot|\bar J^*_c| \geq \frac{\epsilon}{2} \cdot |J^*_c|$.
   
    Notice that the events that $j \in R$ for all $j \in \tilde J^*_{t,c}$ are non-positively correlated. So,
    \begin{align*}
    	\Pr[\bar J^*_{c} \cap R^t_1 = \emptyset] \leq \Pr[\tilde J^*_{t,c} \cap R^t_1 = \emptyset] \leq \left(1 - \frac{\ln(kn)}{|\tilde J^*_{t,c}|}\right)^{|\tilde J^*_{t,c}|} \leq e^{- \ln(kn)} = \frac1{kn}.
    \end{align*}

    Now fix some $c \in [k]$ of type-1b. By the definition of type-1b colors, we have for some $t \in \bfT$ $\sum_{j \in \bar J^*_c}\big((1-\epsilon)b^t_j - t\big)^+ \geq \frac{\epsilon^2}{k}\cdot\sum_{j \in C}\big((1-\epsilon)b^t_j - t\big)^+$. By the way we sample $R^t_2$ in Step 2 of Algorithm~\ref{algo:top-t-cap-choose-R}, we have
    $\Pr[\bar J^*_c \cap R^t_2 = \emptyset] \leq \frac{1}{n^2}$ if the hidden constant is large enough.  The lemma follows from the union bound over the at most $k$ type-1 colors. 
\end{proof}

\subsection{Step~\ref{step:MNCkC-pivots} and \ref{step:MNCkC-radius} of Algorithm~\ref{algo:MNCkC}: Guessing Pivots, Optimum Facilities and Radius}
In Step~\ref{step:MNCkC-pivots} of Algorithm~\ref{algo:MNCkC}, we guess the \emph{pivot} $p_c$ for type-1 or 2 colors $c$, which is defined as follows. 
\begin{itemize}
    \item If $c$ is of type-1, then $p_c$ is defined as any client in $\bar J^*_c \cap R$. $p_c \in R \subseteq C$ in this case.
    \item  If $c$ is of type-2, then $p_c$ is defined as the $i^t_j$ for some $t \in \bfT$ and $ j \in \tilde J^*_{t,c}$ satisfying $u_{i^t_j} \geq k|J^*_c|$ with the smallest $b_j^t$ value. Then $p_c \in S \subseteq F$. Such a $(t, j)$ pair exists, since otherwise $c$ would be of type-1a.
\end{itemize}
We also guess $i^*_c$ for each type-3 color $c$. By definition, we have $i^*_c \in S$ and thus we can afford this.

In Step~\ref{step:MNCkC-radius}, we guess $d(i^*_c, p_c)$ for every type-1 or type-2 color $c$. We can afford to guess a $(1+\epsilon)$-approximation of $d(i^*_c, p_c)$: $r_c$ is the smallest $(1+\epsilon)^z$ that is at least $d(i^*_c, p_c)$ for an integer $z$.

So, after the guessing, we know the $p_c$'s for type-1 and 2 colors $c$, and $i^*_c$'s for type-3 colors $c$. However, we do not know $i^*_c$'s for type-1 and 2 colors $c$. 

\subsection{Step~\ref{step:MNCkC-return} of Algorithm~\ref{algo:MNCkC}: Finding Clusters Using Pivots and Radius}

With all the guessed information,  we can construct a feasible clustering.  This is done in Step~\ref{step:MNCkC-return} of Algorithm~\ref{algo:MNCkC}, which calls $\texttt{MNCkC-clustering-with-pivots}()$ as described in Algorithm \ref{algo:MNCkC-clustering-with-pivots}. 

\begin{algorithm}[h]
    \caption{$\texttt{MNCkC-clustering-with-pivots}()$}
    \label{algo:MNCkC-clustering-with-pivots}
    \begin{algorithmic}[1]
        \State $T \gets \emptyset$
        \For{every color $c \in [k]$ of type-1} 
            \State $q_c \gets \arg\max\big\{u_i: i \in \ball_{F \setminus S}(p_c,r_c) \cap \mathtt{color}^{-1}(c)\big\}$; $T \gets T \cup \{q_c\}$
        \EndFor
        \For{every color $c \in [k]$ of type-3}
            \State $T \gets T \cup \{i^*_c\}$
        \EndFor
        \For{every color $c \in [k]$ of type-2}
            \If{$p_c \neq i^*_{c'}$ for every type-3 color $c'$}
                $T \gets T \cup \{p_c\}$
            \Else
                \ $g_c \gets \arg\max\big\{u_i: i \in \ball_{F \setminus S}(p_c,r_c) \cap \mathtt{color}^{-1}(c)\big\}$; $T \gets T \cup \{g_c\}$
            \EndIf
        \EndFor
        \State \Return $T$ and the feasible assignment $\sigma \in T^C$ respecting the capacity constraints, either directly (if $f$ is the $\topell$ norm) or using Theorem \ref{thm:find-assignment} (if $f$ is general)
    \end{algorithmic}
\end{algorithm}

Notice that for a type-2 color $c$, $p_c$ is not always of color $c$, so Step $8$ in Algorithm \ref{algo:MNCkC-clustering-with-pivots} is not redundant. Further, we treat $T$ as a set, instead of a multi-set, so $|T| \leq k$ and it may happen that $|T| < k$.

\subsection{Analysis of Cost}

% \begin{proof}

%     To prove the lemma, it suffices to show that the $\topnorm{\ell}$ cost of $\phi$ is bounded by $(2+O(\epsilon))$ times the $\topnorm{\ell}$ cost of $\calS^*$. For each client $j$, let $\beta_j$ be its moving distance in this step. The remaining part is dedicated to prove \begin{equation*}
%     \sum_{j \in C}\left(\beta_j - (2+O(\epsilon))t'\right)^+ \le (2+O(\epsilon)) \sum_{c \in [k], j \in J^*_C} (d(i,j)-t')^+.
% \end{equation*}
%     Notice that this inequality directly implies our claim. The proof is essentially the same as the analysis in Algorithm \ref{algo:MNCkC-clustering-with-pivots}, with small change for type-2 colors. For type-2 color $c$, as we choose $p_c$ as $i_{j,l_0}$ with the least $b_{j,l_0} = d(p_c,i_{j,l_0})$ values, further as $c$ is not type-1a, there exists $p_c' = i_{j',t}$ with $u_{i_{j',t}} \ge k|J^*_c|.$ So we have $b_{p_c'} \ge b_{p_c}$, and we can proceed with the same analysis. 
% \end{proof}

To analyze the cost of the solution, we explicitly construct an assignment for $T$ with small cost. We build a bipartite graph $H = ([k], T, E_H)$ as follows; it is instructive to correlate the construction with Algorithm~\ref{algo:MNCkC-clustering-with-pivots}.
\begin{itemize}
    \item For every $c$ of type-1, we add $(c,q_c)$ to $E_H$.
    \item For every $c'$ of type-3 (which implies $i^*_{c'} \in T$), we add $(c', i^*_{c'})$ to $E_H$.
    \item For every $c$ of type-2, we add $(c, p_c)$ to $E_H$. If additionally $p_c = i^*_{c'}$ for some type-3 color $c'$, then we add $(c', g_c)$ to $E_H$.
\end{itemize}

We construct a solution in the following way:
Initially, all clients are moved to their corresponding facilities in $\calS^*$, with moving cost precisely $\opt$. After that, $i^*_c$ contains $|J^*_c| \leq u_{i^*_c}$ clients for any color $c$.
Then, we move the clients from $S^*$ to $T$ according to a ``transportation'' function $\phi \in \Z_{\geq 0}^{E_H}$: $\phi_{c, i}$ clients will be moved from $i^*_c$ to $i$. In order for $\phi$ to be a solution, $\phi$ must satisfy the following properties:
    \begin{itemize}
        \item $\forall c \in [k], \sum_{(c, i)\in E_H} \phi_{c, i} = |J^*_c|$;
        \item$\forall i \in T, \sum_{(c, i)\in E_H} \phi_{c, i} \leq u_i$. 
    \end{itemize}

\begin{figure}
    \centering
    \includegraphics[width=0.45\linewidth]{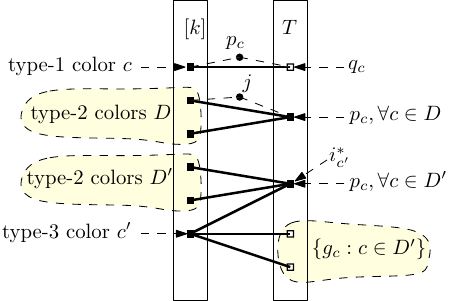}
    \caption{3 types of connected components of $H$. In the first type, a type-1 color $c$ is connected to $q_c$. Notice that $q_c$ is defined via a representative client $p_c \in R \cap \bar J^*_c$. In the second type, we have many type-2 colors $c \in D$ connected to their common $p_c$. $p_c$ for a type-2 color $c$ is defined via a client $j$. In the third type, we have many type-2 colors $c \in D$ ($D'$ in the figure to avoid confusion) connected to their common $p_c$, which is $i^*_{c'}$ for a type-3 color $c'$. Then we have edges $(c', i^*_{c'})$ and $(c', g_c)$ for every $c \in D$ in the component. Solid squares in $T$ are in $S$, and empty squares in $T$ are not in $S$.}
    \label{fig:H}
\end{figure}

In the following, we construct $\phi$. Connected components in $H$ are only of the following three possibilities, and we construct $\phi$ for each of them (See Figure~\ref{fig:H} for an illustration):
\begin{enumerate}[(1)]
    \item Some type-1 color $c$ connected to $q_c \in T$. Then $q_c$ has degree $1$ and $\{c, q_c\}$ is maximally connected. We set $\phi_{c, q_c} = |J^*_c|$. %As $i^*_c$ is a candidate for $q_c$, 
    We have $u_{q_c} \geq u_{i^*_c} \geq |J^*_c|$. 
    
    \item Several type-2 colors and no type-3 color connect to facility $i \in T$. Let $D:=\{c:p_c = i\}$, then $D \cup \{i\}$ is maximally connected. Set $\phi_{c, i} = |J^*_c|$ for every $c \in D$. We have $\sum_{c \in D}|J^*_c| \leq \sum_{c \in D}\frac{u_i}{k} \leq u_i$ as all colors in $D$ are of type-2. 

    \item Facility $i = i^*_{c'} \in T \cap S^*$ is connected to one type-3 color $c'$. Some type-2 colors may also connect to $i$. Let $D:=\{c:p_c = i, c \text{ is of type-2}\}$; $D$ may be empty. $\{c',i\} \cup D \cup \{g_c \mid c \in D\}$ is one connected component in $H$. Similar to (ii), $\sum_{c \in D}|J^*_c| \leq u_i$. We define $\phi_{c, i} = |J^*_c|$ for every $c \in D$. We then define $\phi_{c', i} = \min\{|J^*_{c'}|, u_i - \sum_{c \in D}|J^*_c|\}$. If $u_i - \sum_{c \in D}|J^*_c| < |J^*_{c'}|$, then we define $\phi_{c', g_c}$ values for all $c \in D$ so that $\phi_{c', g_c} \leq |J^*_c|$ for every $c \in D$, and $\sum_{c \in D}\phi_{c', g_c} = |J^*_{c'}| - \phi_{c', i}$. This is feasible since $\sum_{c \in D} |J^*_c| = u_i - \phi_{c', i} \geq |J^*_{c'}| - \phi_{c', i}$. As for $c \in D$, $i^*_c$ is a candidate for $g_c$, $u_{g_c} \ge u_{i_c^*} \ge |J_c^*|$, so $\phi$ does not violate the capacity constraint for $g_c$.
\end{enumerate}
In summary, the two properties needed for the transportation function $\phi$ are satisfied, which implies the assignment of $C$ to $T$ respects all capacity constraints. 

We then bound the $f$-norm cost of moving clients from $S^*$ to $T$ using the transportation function $\phi$. For each client $j$, let $\beta_j$ be the moving distance of this step, and $b^*_j$ be the connection distance of $j$ in the optimum solution $\calS^*$. The main lemma we prove is the following:
\begin{lemma}
    \label{lemma:bound-for-each-t}
    For every $t \in \bfT$, we have
    \begin{equation*}
        \sum_{j \in C}\left(\beta_j - (2+O(\epsilon))t\right)^+ \le (2+O(\epsilon)) \sum_{j \in C} (b^*_j-t)^+.
    \end{equation*} 
\end{lemma}

\begin{proof}
    Throughout the proof, we fix a $t \in \bfT$.  We distinguish different color types to prove the inequality.

    For a type-1 color $c$, $J^*_c$ will be moved from $i^*_c$ to $q_c$. 
    \begin{align*}
        d(i^*_c, q_c)\leq d(i^*_c, p_c) + d(p_c, q_c) \leq (2+\epsilon) d(i^*_c, p_c) \leq \frac{2+\epsilon}{1-\epsilon}\cdot \frac{1}{|J^*_c|} \sum_{j \in J^*_c} d(i^*_c, j) = \frac{2+\epsilon}{1-\epsilon}\cdot \frac{1}{|J^*_c|} \sum_{j \in J^*_c} b^*_j.
    \end{align*}
    The second inequality is by that $r_c \leq (1+\epsilon)d(i^*_c, p_c)$ and $q_c \in \ball_{F \setminus S}(p_c, r_c)$. The third inequality is by that $p_c \in \bar J_c^*$ and $\bar J^*_c$ are the $\ceil{\epsilon|J^*_c|}$ clients $j$ in $J^*_c$ with the smallest $d(i^*_c, j)$ values. 
    So, 
    \begin{align}
        &\quad \sum_{j \in J_c^*} \Big(\beta_j - \frac{2+\epsilon}{1-\epsilon}\cdot t\Big)^+ 
        =|J^*_c|\cdot\Big(d(i^*_c, q_c) - \frac{2+\epsilon}{1-\epsilon}\cdot t\Big)^+ 
        =\Big(|J^*_c|\cdot d(i^*_c, q_c) - \frac{2+\epsilon}{1-\epsilon}\cdot |J^*_c| \cdot t\Big)^+ \nonumber\\
        &\leq \Big(\frac{2+\epsilon}{1-\epsilon}\cdot\sum_{j \in J^*_c} b^*_j - \frac{2+\epsilon}{1-\epsilon}\cdot |J^*_c| \cdot t\Big)^+ \leq \frac{2+\epsilon}{1-\epsilon} \cdot \sum_{j \in J^*_c} (b^*_j - t)^+ = (2+O(\epsilon))\sum_{j \in J^*_c} (b^*_j - t)^+. \label{inequ:bound-type1-final}
    \end{align}

    We consider type-2 and type-3 colors together. For the case (ii) above, we moved $|J^*_c|$ clients from $i^*_c$ to $i$ for every $c \in D$. For the case (iii), we did the same; additionally, for every $c \in D$, we moved at most $|J^*_c|$ clients from $i^*_{c'} = p_c$ to $g_c$. Notice that by the definition of $g_c$ for every $c \in D$, we have $d(i^*_{c'}, g_c) = d(p_c, g_c) \leq r_c \leq (1+\epsilon) d(p_c, i^*_c)$. Some clients in $i^*_{c'}$ stayed at $i = i^*_{c'}$. Therefore, we have
    \begin{align}
        \sum_{c\text{ type-2 or 3}, j \in J^*_c} \Big(\beta_j - \frac{(1+\epsilon)(2-\epsilon)}{1-\epsilon} \cdot t\Big)^+ &\leq 2 \sum_{c\text{ type-2}}|J^*_c|\cdot\Big((1+\epsilon)\cdot d(i^*_c, p_c) - \frac{(1+\epsilon)(2-\epsilon)}{1-\epsilon} \cdot t\Big)^+ \nonumber\\
        &=2(1+\epsilon) \sum_{c\text{ type-2}}|J^*_c|\cdot\Big(d(i^*_c, p_c) - \frac{2-\epsilon}{1-\epsilon}\cdot t\Big)^+. \label{inequ:bound-type23}
    \end{align}

    Now, we focus on each type-2 color $c$. By the definition of $p_c$, we have $p_c = i^t_{j'}$ for some $j' \in \tilde J^*_{t,c}$. As $c$ is not of type 1, $\sum_{j \in \bar J^*_c}((1-\epsilon)b^t_j - t)^+ < \frac{\epsilon^2}{k}\cdot \sum_{j \in C}((1-\epsilon)b^t_j-t)^+$. We have 
    \begin{align*}
        \Big(d(j', p_c) - \frac1{1-\epsilon}\cdot t\Big)^+ &= \Big(b^t_{j'} - \frac1{1-\epsilon}\cdot t\Big)^+ \leq \frac{2}{|\bar J^*|} \sum_{j \in \bar J^*_c}\Big(b^t_j - \frac1{1-\epsilon}\cdot t\Big)^+ \\
        &= \frac{2}{(1-\epsilon)|\bar J^*_c|} \sum_{j \in \bar J^*_c}\big((1-\epsilon)b^t_j - t\big)^+  \leq \frac{2\epsilon^2}{(1-\epsilon)k|\bar J_c^*|}\cdot\sum_{j \in C}\big((1-\epsilon)b^t_j - t\big)^+ \\
        &\leq \frac{2\epsilon}{(1-\epsilon)k|J_c^*|} \sum_{j \in C} \big((1-\epsilon)b^t_j - t\big)^+ = \frac{O(\epsilon)}{k|J_c^*|} \sum_{j \in C} \big((1-\epsilon)b^t_j - t\big)^+.
    \end{align*}
    The first inequality used that $j' \in \tilde J^*_{t, c}$ and $\tilde J^*_{t, c}$ contains the $\ceil{\frac{|\bar J^*_c|}{2}}$ clients in $\bar J^*_c$ with the smallest $b^t_j$ values. The third inequality used that $|\bar J^*_c| \geq \epsilon |J^*_c|$. 

    As $j' \in  \tilde J^*_{t,c} \subseteq \bar J^*_c$, we have 
    \begin{align*}
        (d(i^*_c, j') - t)^+ \leq \frac{1}{1-\epsilon} \cdot \frac{1}{|J^*_c|}\sum_{j \in J^*_c} (d(i^*_c, j) - t)^+ = \frac{1+O(\epsilon)}{|J^*_c|}\sum_{j \in J^*_c} (b^*_j - t)^+.
    \end{align*}
    
    Combining the above two inequalities, we have 
    \begin{align*}
        |J^*_c|\cdot\Big(d(i^*_c, p_c) - \frac{2-\epsilon}{1-\epsilon}\cdot t\Big)^+ &\leq |J_c^*|\cdot \Big(d(i^*_c, j')-t\Big)^+ + |J_c^*|\cdot \Big(d(j', p_c)-\frac{1}{1-\epsilon}\cdot t\Big)^+ 
        \\ 
        &\leq  (1+O(\epsilon)) \sum_{j \in J^*_c} (b^*_j-t)^+ + \frac{O(\epsilon)}{k}\sum_{j \in C}((1-\epsilon)b^t_j - t)^+. 
    \end{align*}

    Applying the upper bound to \eqref{inequ:bound-type23}, we get
    \begin{align}
        &\quad\sum_{c\text{ type-2 or 3}, j \in J^*_c} \Big(\beta_j - \frac{(1+\epsilon)(2-\epsilon)}{1-\epsilon}\cdot t\Big)^+ \nonumber\\
        & \leq (2+O(\epsilon))\cdot \sum_{c\text{ type 2}}\sum_{j \in J^*_c} (b^*_j - t)^+ +  O(\epsilon) \cdot \sum_{j \in C} \big((1-\epsilon)b^t_j - t\big)^+. \label{inequ:bound-type23-final}
    \end{align}
    
    Summing up contribution for colors all the 3 types, by adding \eqref{inequ:bound-type1-final} and \eqref{inequ:bound-type23-final}, we get 
    \begin{align*}
        &\quad \sum_{j \in C}\left(\beta_j - (2+O(\epsilon))t\right)^+ = \sum_{c \in [k]} \sum_{j \in J^*_c} (\beta_j - (2+O(\epsilon))t)^+
        \\ &\le  (2+O(\epsilon)) \sum_{j \in C} (b^*_j-t)^+ + O(\epsilon) \sum_{j \in C} ((1-\epsilon)b^t_j-t)^+ 
        \le (2+O(\epsilon)) \sum_{j \in C} (b^*_j-t)^+.
    \end{align*}
    The last inequality used $\sum_{j \in C} ((1-\epsilon)b^t_j-t)^+ \le \sum_{j \in C} (b^*_j-t)^+$, which is from Property (P).
    \end{proof}

    Recall that in our constructed solution, we moved clients from $C$ to $T$ in two steps. We first moved $C$ to $S^*$ according to $\calS^*$, which incur a cost of $\opt$. Then we moved clients from $S^*$ to $T$, with moving distance vector $\beta$. 

    Now we describe the $f \equiv \topell$ and general $f$ case separately. For the case $f \equiv \topell$, Lemma~\ref{lemma:bound-for-each-t} holds for the $\ell$-th largest coordinate of $b^*$. By Lemma \ref{lemma:linearize-top-l}, the moving cost of the second step is 
    \begin{align*}
        |\beta|_{\topell} &\leq \ell \cdot (2+O(\epsilon)) t + \sum_{j \in C}(\beta_j - (2+O(\epsilon))t)^+ \leq (2+O(\epsilon))\ell t + (2+O(\epsilon)) \sum_{j \in C}(b^*_j - t)^+ \\
        &\leq (2+O(\epsilon))\left(\ell t + \sum_{j \in C}(b^*_j - t)^+\right) = (2+O(\epsilon))\cdot |b^*|_\topell = (2+O(\epsilon))\opt.
    \end{align*}
    This implies that $\topell$ cost of our constructed solution is at most $(3+O(\epsilon))\cdot \opt$. 

    Now consider a general monotone symmetric norm $f$. We show that for every integer $\ell \in [n]$, we have $|\beta|_\topell \leq (2+O(\epsilon)) |b^*|_\topell$. By Lemma~\ref{lemma:symmetric-to-top}, this implies $|\beta|_f \leq (2+O(\epsilon)) |b^*|_f = (2+O(\epsilon))\opt$. Focus on any $\ell \in [n]$ and let $t$ be the $\ell$-th largest coordinate in $b^*$. Then, we have $t' = (1+\epsilon)^{\ceil{\log_{1+\epsilon}t}} \in \bfT$. Applying Lemma~\ref{lemma:bound-for-each-t} with $t$ being this $t'$ will prove $|\beta|_\topell \leq (2+O(\epsilon)) |b^*|_\topell$. Therefore, the $f$-norm cost our constructed solution is at most $(3+O(\epsilon))\cdot \opt$. In both cases, scaling $\epsilon$ at the beginning gives us a $(3+\epsilon)$-approximation.

\subsection{Analysis of Runtime}
Now we analyze the running time of our algorithm. When all guesses are correct, the coloring satisfies our requirement, and the events mentioned in Lemma \ref{lem:existence-ra} happen, the algorithm returns a solution with value at most $(3+O(\epsilon)) \opt$.

For the runtime, all steps except coloring and guessing are polynomial. With probability $\frac{k!}{k^k}$, the color function in Step~\ref{step:MNCkC-color} satisfies our requirement, so we need to run the algorithm $O(k^k \ln n)$ times to boost the success probability to $1-\frac{1}{n}$. For the guessing part, Step~\ref{step:MNCkC-type} has $3^k$ possible choices for the types, Step~\ref{step:MNCkC-pivots} has $(|R| + |S|)^k = (\frac{k \ln n}{\epsilon})^{O(k)}$ choices for pivots, and finally, as we can always assume $d(c,f) = \mathrm{poly}(n)$ for all $c \in C$ and $f \in F$, Step~\ref{step:MNCkC-radius} has $(\log_{1+\epsilon} \mathrm{poly}(n))^k = \left(\frac{\ln n}{\epsilon}\right)^{O(k)}$ choices for the radius. In total, the running time is $(\frac{k \ln n}{\epsilon})^{O(k)} n^{O(1)}$, which can be bounded by $(\frac{k}{\epsilon})^{O(k)} n^{O(1)}$. \footnote{We need to bound $(\ln n)^{O(k)}$. If $k \leq \sqrt{\log n}$, then this is upper bounded by $n^{O(1)}$. Otherwise, it is upper bounded by $k^{O(k)}$.}
    \section{FPT Time $(1 + \frac2{ec} + \epsilon)$-Approximation for Top-$cn$ Norm $k$-Clustering} 
\label{sec:top-cn-kC}
In this section, we give the tight $(1 + \frac2{ec} + \epsilon)$-approximation algorithm for the $\topcn$ norm $k$-clustering problem for $c \in (\frac1e, 1]$, proving Theorem~\ref{theo: top-cn norm}.  Notice that the case $c \leq \frac1e$ has an approximation ratio of $3+\epsilon$, as stated in Corollary~\ref{coro:three-approximation}.

\subsection{Useful Tools for Top-$\ell$ Norms}
            
    In this section, we describe some useful tools. We assume the distances in the metric $d$ are non-negative integers. It will be convenient to use the \emph{occurrence-times vectors} to represent distance vectors: for each $a \in \Z_{\geq 0}$, we have a coordinate indicating the number of times $a$ appear in the distance vector; that is, how many clients have distance $a$ to their nearest facility.  With this in mind, we define the set of occurrence-times vectors to be
			\begin{align*}
				\calS_n := \{\delta \in \R_{\geq 0}^{\Z_{\geq 0}}: |\delta|_1 = n\text{ and } \delta\text{ has a finite support}\}. 
			\end{align*} 
			For a $\delta \in \calS_n$ and $a \in \Z_{\geq 0}$, we can think of $\delta_a$ as the number of times $a$ appears in the distance vector.  It is convenient to allow fractional occurrence times.  

            For any vector $\phi \in \R_{\geq 0}^{Z\geq 0}$ with finite support, we define $\overline{L_1}(\phi) = \sum_{a = 0}^\infty \phi_a \cdot a$. (Throughout, we shall typically use $\overline{f}$ to denote a norm function $f$ using the occurrence-times vector representation{:  for an occurrence-time vector $\delta\in \calS_n$, $\bar f(\delta)$ is the value of $f$ on the corresponding multiset of distances.)

			\begin{definition}
				\label{def:top-l-occur}
				For every $\delta \in \calS_n$ and real $\ell \in [0, n]$, we define $\overline{\topell}(\delta)$ to be the value of the following linear program with variables $\alpha \in \R_{\geq 0}^{\Z_{\geq 0}}$: maximize $\overline{L_1}(\alpha)$ subject to $0 \leq \alpha \leq \delta$ and $|\alpha|_1 = \ell$.
			\end{definition}
			To get some intuition about the definition, consider the case where $d$ is a distance vector of dimension $n$ and $\delta$ is its correspondent occurrence-times vector. Then $\overline{\topell}(\delta) = |d|_{\topell}$. We extended the definition to real vectors $\delta$.
									
			\begin{restatable}{lemma}{lemmaconcave}
                \label{lemma:concave}
				For every real $\ell \in [0, n]$, we have that $\overline{\topell}(\cdot)$ is concave on $\calS_n$. 
			\end{restatable}
			
			This is in contrast to the $\topell$ norm function using the normal representation of vectors, which is convex. 
			
		\begin{definition}
			We say a vector $\delta \in \calS_n$ dominates a vector $\delta' \in \calS_n$ with a factor of $\gamma$ for some real $\gamma \geq 0$, denoted as $\delta' \preceq_\gamma \delta$, if $ \overline{\topell}(\delta')\leq \gamma\cdot\overline{\topell}(\delta)$ for every real $\ell \in [0, n]$. We use $\preceq$ for $\preceq_1$. 
		\end{definition}

        \begin{restatable}{lemma}{lemmadominatingaddingtransitive}
            \label{lemma:dominating-adding-transitive}
            Assume $\delta^1, \delta^2, \cdots, \delta^H, \delta'^1, \delta'^2, \cdots, \delta'^H$ are $2H$ vectors in $\calS_n$, $\gamma \in \R_{>0}$, and $\delta'^h \preceq_\gamma \delta^h$ for every $h \in [H]$. Let $\beta_1, \beta_2, \cdots, \beta_H \in [0, 1]$ satisfy $\sum_{h = 1}^H \beta_h = 1$. Then
            \begin{align*}
                \sum_{h = 1}^H \beta_h\delta^{'h} \preceq_\gamma \sum_{h = 1}^H \beta_h\delta^{h}. 
            \end{align*}
        \end{restatable}

		\begin{restatable}{corollary}{corodominatingaveraging}
        \label{coro:dominating-averaging}
			Let $\delta \in \calS_n$, and $\delta = \sum_{h = 0}^H \theta^h$ where $H \in \Z_{\geq 0}$ and $\theta^0, \theta^1, \cdots, \theta^H \geq 0$. Let $\gamma \geq 1$. For every $h \in [H]$, let $b_h \in \Z_{\geq 0}$ satisfy $b_h \leq \gamma \cdot \frac{\overline{L_1}(\theta^h)}{|\theta^h|_1}$. Let $\delta' = \theta^0 + \sum_{h=1}^H|\theta^h|_1\cdot {\bf e}_{b_h}$, where ${\bf e}_b$ for any $b \in \Z_{\geq 0}$ is the vector in $\R_{\geq 0}^{Z_{\geq 0}}$ with $({\bf e}_b)_b = 1$ and $({\bf e}_b)_{b'} = 0$ if $b' \neq b$. 
            
            Then $\delta' \preceq_\gamma \delta$.
		\end{restatable}

        We treat $\delta$ as $n$ fractional values. Each $\theta^h$ contains a disjoint portion of $\delta$. For every $h \in H$, we replace $\theta^h$ with $|\theta^h|_1$ fractional values equaling to $b_h$. The lemma says that the resulting occurrence-times vector is dominated by the original one with a factor of $\gamma$.

        \begin{definition}
            \label{def:adding-occurrence-vectors}
            For any two vectors $\delta, \delta' \in \calS_n$, we say some $\phi \in \calS_n$ can be obtained by adding $\delta$ and $\delta'$ if there exists some $z \in \R_{\geq 0}^{\Z_{\geq 0} \times \Z_{\geq 0}}$ such that 
			\begin{itemize}
				\item $\sum_{a'}z_{a, a'} = \delta_a$ for every $a \in \Z_{\geq 0}$,
				\item $\sum_{a}z_{a, a'} = \delta'_{a'}$ for every $a' \in \Z_{\geq 0}$, and
                \item $\phi_t = \sum_{a + a' = t}z_{a, a'}$ for every $t \in \Z_{\geq 0}$.
			\end{itemize}
        \end{definition}
        $z$ gives a matching between the vectors $\delta$ and $\delta'$. Then the occurrence-time vector $\phi$ is obtained by adding the distance vectors for $\delta$ and $\delta'$, using the matching $z$. 

        \begin{definition}
            Let $\delta, \delta' \in \calS_n$, and $z \in \R_{\geq 0}^{\Z_{\geq 0} \times \Z_{\geq 0}}$ be the unique vector satisfying the first two conditions in Definition~\ref{def:adding-occurrence-vectors} and the following condition:
            \begin{itemize}
                \item There are no integers $a < a', b > b'$ such that $z_{a, b} > 0$ and $z_{a', b'} > 0$. 
            \end{itemize}
            Then, we use $\delta \oplus \delta'$ to denote the vector $\phi \in \calS_n$ satisfying the third condition of Definition \ref{def:adding-occurrence-vectors}. We use $\delta \otimes 2$ to denote $\delta \oplus \delta$.
        \end{definition}

        \begin{lemma}
            \label{lemma:dominating-for-adding}
            Let $\delta, \delta' \in \calS_n$, $\phi = \delta \oplus \delta'$, $\phi' \in \calS_n$ be obtained by adding $\delta$ and $\delta'$. Then $\phi' \preceq \phi$.
        \end{lemma}

 %The main theorem we need to use is the following:
		\begin{restatable}{theorem}{thmmixoccurrence}
			\label{thm:mix-occurrence}
			Let $\delta, \delta' \in \calS_n$ such that $\delta' \preceq_\gamma \delta$ for some $\gamma > 0$. Let $c \in (\frac1e ,1]$ and $\alpha \in [0, c]$.  Then we have 
			\begin{align*}
				\overline\topcn\big((1-\alpha) \delta + \alpha\cdot(\delta \oplus \delta'\otimes 2)\big) \leq \left(1 + \frac{2 \alpha\gamma}{c}\right)\overline\topcn(\delta).
			\end{align*}
		\end{restatable}
		
		We only apply the theorem for $\alpha = \frac1e$ and $\gamma = 1 + O(\epsilon)$.

%For the ($k$-center, $k$-median) objective, the ratio is $(3, 1+\frac2e)$.  This already improves upon the $(4, 8)$ approximation ratio proposed by Alamdari and Shmoys cite{Alamdari2017} that is achieved by polynomial-time algorithms. 
%
%For ease of extending the algorithm to other norms, we describe a generic framework defined in Algorithm~\ref{algo:generic}. Throughout this section, we fix a small enough constant $\epsilon > 0$.  

\subsection{The Algorithm}
       Throughout this and the next section, given any set $S \subseteq F$ (it is possible that $|S| \neq k$), we shall use $\vec{d}(S)$ to denote the vector $(d(j, S))_{j \in C}$ of $n$ distances; we call it the distance vector for $S$.  So the goal of the $\topcn$ norm $k$-clustering problem is to minimize $|\vec d(S)|_\topcn$ subject to $S \subseteq F, |S| = k$.

Let $S^*$ be the unknown set of $k$ facilities in the optimum solution.  We let $\opt = |\vec d(S^*)|_\topcn$.  For every $i^* \in S^*$, and the set $J \subseteq C$ of clients connected to $i^*$ in the optimum solution, we refer to $(i^*, J)$ as an \emph{optimum cluster}.  
\begin{definition}[Per-Client Costs  and Cores]
    For every optimum cluster $(i^*, J)$, we define its \emph{per-client cost}  to be  $\frac{1}{|J|}\sum_{j\in J}d(i^*, j)$.  We define its \emph{core} to be the set of the $\ceil{\epsilon|J|}$ clients in $J$ with the smallest $d(i^*, j)$ values.
\end{definition}

%TODO: we are not trying to optimize the dependence of the running time on $k$ and $\epsilon$. There are many places where more involved careful analysis can lead to a better dependence. 

\begin{algorithm}[h]
	\caption{Top-$cn$ Norm $k$-Clustering} 
	\label{algo:top-cn}
	\begin{algorithmic}[1]
		\State obtain a set $\calS$ of valid stars using Theorem~\ref{thm:pseudo-approx} with $h\equiv\topcn$ and $u_i=\infty,\forall i\in F$
        \State $S \gets \{i \in F: \exists J, (i, J) \in \calS\}$ \label{step:Tpcn-calS}
        %obtain a $(1+\epsilon)$-approximate solution $S \subseteq F$ of size $O(\frac{k\log n}{\epsilon})$ using LP rounding:
        \label{step:Tpcn-S}
		%\State sample a set $R \subseteq C$ of at most $k$ clients, that we call \emph{representatives}  
     	\State $R \gets \texttt{TpcnC-choose-R}()$
         \Comment{Clients in $R$ are called \emph{representatives}} \label{step:Tpcn-R}
		\State guess a multi-set $P \subseteq S \cup R$ of size $k$ \label{step:Tpcn-P} \Comment{Points in $P$ are called \emph{pivots}}
        \State guess a vector $(r_p)_{p\in P}$ where each $r_p$ is $0$ or an integer power of $(1 + \epsilon)$ in $[1, \Delta]$ \footnotemark  \label{step:Tpcn-r}
        \State \Return $\texttt{TpcnC-clustering-with-pivots}()$ \label{step:Tpcn-return}
  \end{algorithmic}
\end{algorithm}
\footnotetext{ Let $\Delta: = \max\{d(u, v): u, v\in F\cup C, d(u, v)< \infty\}$. Recall that after the standard discretization, all finite distances are integers in $[0, \Delta]$, with $\Delta =\poly(n)$.}

The pseudo-code for the algorithm is given in Algorithm~\ref{algo:top-cn}. In Step~\ref{step:Tpcn-calS}, we apply Theorem~\ref{thm:pseudo-approx} to obtain a solution $\calS$ and let $S$ be the set of open facilities. As there are no capacities, we simply use $S$ to denote the solution and discard the notion $\calS$. We have $|S| \leq \frac{O(k \log n)}{\epsilon}$. Moreover, $|\vec d(S)|_\topcn \leq \frac{\opt}{1-\epsilon}$.

We describe the remaining steps of the algorithm in more detail.

		\subsection{Step~\ref{step:Tpcn-R} of Algorithm~\ref{algo:top-cn}: Choosing Representatives}

        \begin{algorithm}[h]
    \caption{$\texttt{TpcnC-choose-R}()$}\label{alg:choose-R-top-cn}
    \begin{algorithmic}[1]
        \State $R \gets \emptyset$
        \State \textbf{repeat} $k$ times:
            \State \hspace*{\algorithmicindent} randomly choose a client $j\in C$, with probabilities proportional to $d(j, S)$
            \State \hspace*{\algorithmicindent} $R \gets R \cup \{j\}$
        \State \Return $R$
    \end{algorithmic}
\end{algorithm}

		In this step, we call the procedure \texttt{TpcnC-choose-R}(), described in Algorithm~\ref{alg:choose-R-top-cn}. Before analyzing the properties of the representative set $R$, we make some definitions and partition the optimum clusters into 3 types: Given an optimum cluster $(i^*, J)$ with core $J'$, we say $(i^*, J)$ is of 
           \begin{itemize}
	           \item \emph{type-1} if $\sum_{j\in J'}d(j, S)\geq \frac{\epsilon^3}{k} \cdot |\vec{d}(S)|_1$,
    	       \item \emph{type-2} if $\sum_{j\in J'}d(j, S)< \frac{\epsilon^3}{k}\cdot |\vec{d}(S)|_1$  and $\sum_{j\in J'}d(i^*, j)\geq \frac{\epsilon^2}{k}\cdot  |\vec{d}(S)|_1$, and 
    	       \item \emph{type-3} if $\sum_{j\in J'}d(j, S)< \frac{\epsilon^3}{k}\cdot |\vec{d}(S)|_1$  and $\sum_{j\in J'}d(i^*, j)< \frac{\epsilon^2}{k} \cdot |\vec{d}(S)|_1$.
           \end{itemize}
    Notice that the three types partition all the optimum clusters. 

\begin{lemma}
    \label{lemma:hit-cores}
    With probability at least $\frac{\epsilon^{3k}}{k^k}$, the following event happens:   For every optimum cluster $(i^*, J)$ of type-1 with core $J'$, we have $J' \cap R \neq \emptyset$. In other words, $R$ intersects the core of every type-1 optimum cluster. 
\end{lemma}
\begin{proof}
	Focus on the process of choosing one $j$ with probability proportional to $d(j, S)$. For the core $J'$ of a type-1 optimum cluster $(i^*, J)$, we have 
	\begin{align*}
		\Pr[j \in J']&=\frac{\sum_{j'\in J'}d(j', 
		S)}{\sum_{j' \in C}d(j', S)} \geq  \frac{\epsilon^3}{k},
	\end{align*}
	%If this happens, we say the client $j$ hits the optimum cluster $(i^*, J)$.
	
	Let $k' \leq k$ be the number of type-1 optimum clusters. We focus on the first $k'$ iterations of the repeat loop of Algorithm~\ref{alg:choose-R-top-cn}, and use $j_1, j_2, \cdots, j_{k'}$ to denote the clients chosen in the $k'$ iterations. Fix any permutation $(i^*_1, J_1), (i^*_2, J_2), \cdots, (i^*_{k'}, J_{k'})$ of the $k'$ clusters. The probability that $R$ hits the cores of all type-1 optimum clusters is at least
	\begin{flalign*}
		&&\Pr[\forall o \in [k']: j_o\text{ is in core of } (i^*_{o}, A_{o})] \geq \left(\frac{\epsilon^3}{k}\right)^{k'} \geq \left(\frac{\epsilon^3}{k}\right)^{k} = \frac{\epsilon^{3k}}{k^k}.&&\qedhere
	\end{flalign*}
\end{proof}

From now on, we assume Step \ref{step:Tpcn-R} of Algorithm~\ref{algo:top-cn} is successful, which means the event in Lemma~\ref{lemma:hit-cores} occurs.

\begin{lemma} \label{lemma:d-to-R-cup-S-small}
 For every type-1 or 2 optimum cluster $(i^*, J)$ with per-client cost $\bar d$, we have $d(i^*, R \cup S) \leq \frac{\bar d}{1-\epsilon}$. 
\end{lemma}
\begin{proof}
	Let $J'$ be the core of $(i^*, J)$. 
	First, assume $(i^*, J)$ is of type-1. As Step~\ref{step:Tpcn-R} is successful, we have $J'\cap R\neq \emptyset$.  Let $p$ be any client in $J' \cap R$. For any $j\in (J\setminus J')\cup \{p\} $, we have $d(i^*, j)\geq d(i^*, p)$. So
      \begin{align*}
          |J|\bar{d}=\sum_{j\in J}d(j, i^*)\geq \sum_{j\in (J\setminus J')\cup \{p\}}d(j, i^*)\geq (|J|-\ceil{\epsilon|J|} + 1)d(i^*, p) \geq (|J|-\epsilon|J|)d(i^*, p).
      \end{align*}
Therefore, we have  $d(i^*, p)\leq \frac{|J|\bar{d}}{|J|-\epsilon|J|}= \frac{\bar d}{1-\epsilon}$.

Then, assume $(i^*, J)$ is of type-2.  By the definition of type-2, we have $\sum_{j \in J'} d(j, S) < \epsilon\sum_{j \in J'}d(i^*, j)$.  Consider the facility $p \in S$ that is closest to $i^*$.  
\begin{align*}
	\sum_{j \in J'}d(i^*, p) \leq \sum_{j \in J'}\big(d(i^*, j) + d(j, S)\big)= \sum_{j \in J'}d(i^*, j)  + \sum_{j \in J'} d(j, S) \leq (1+\epsilon)\sum_{j \in J'}d(i^*, j).
\end{align*}
So, 
\begin{align*}
	d(i^*, p) \leq (1+\epsilon)\cdot\frac{\sum_{j \in J'}d(i^*, j)}{|J'|} \leq (1+\epsilon)\cdot\frac{\sum_{j \in J}d(i^*, j)}{|J|} = (1+\epsilon)\bar d \leq \frac{\bar d}{1-\epsilon}.
\end{align*}
The second inequality used that $d(i^*, j) \leq d(i^*, j')$ for every $j \in J'$ and $j' \in J \setminus J'$. 
\end{proof}

\subsection{Steps~\ref{step:Tpcn-P} and \ref{step:Tpcn-r} of Algorithm~\ref{algo:top-cn}: Guessing Pivots and Radius Vector}
In Steps~\ref{step:Tpcn-P} and \ref{step:Tpcn-r} of Algorithm~\ref{algo:top-cn}, we guess a multi-set $P$ of $k$ pivots from $R \cup S$, and a radius $r_p$  for every $p \in P$. There is a \emph{desired} pivot $p$ for every optimum cluster $(i^*, J)$, and the pivot $p$ has a \emph{desired} radius. Step~\ref{step:Tpcn-P} is successful if $P$ is the set of $k$ desired pivots for the $k$ optimum clusters, and Step~\ref{step:Tpcn-r} is successful if $r_p$ for each $p \in P$ is the desired radius for $p$. For every optimum cluster $(i^*, J)$, the desired pivot $p$ for the cluster and its desired radius are defined as follows:
\begin{itemize}
	\item If $(i^*, J)$ is of type-1 or 2, then $d(i^*, S \cup R) \leq \frac{\bar d}{1-\epsilon}$ by Lemma~\ref{lemma:d-to-R-cup-S-small}. The desired pivot $p$ is the closest point in $S \cup R$ to $i^*$. Thus, we have $d(i^*, p) \leq \frac{\bar d}{1-\epsilon}$.  The desired radius is $d(i^*, p)$  rounded up to the nearest integer power of $1+\epsilon$. Therefore, if $r_p$ is the desired radius, we have $r_p \leq \frac{1+\epsilon}{1-\epsilon}\bar d$.
	\item If $(i^*, J)$ is of type-3, then the desired pivot $p$ is the closest facility in $S$ to $i^*$, and its desired radius is $0$. 
\end{itemize}
Again, we assume the two steps are successful from now on. Therefore, $P$ is the set of desired pivots for the optimum clusters and $r_p$ for each $p \in P$ is the desired radius for $p$. 
\begin{lemma}
    \label{lemma:desired-P-radius}
    There is a set $S'^*$ of $k$ facilities, one from $\ball_F(p, r_p)$ for each $p \in P$, and an assignment vector $\sigma \in (S'^*)^C$ such that the following properties hold. 
    \begin{enumerate}[(\ref{lemma:desired-P-radius}a)]
        \item $\Big|\big(d(j, \sigma_j)\big)_{j \in C}\Big|_\topcn\leq(1+O(\epsilon))\cdot\opt$.  \label{property:sigma-good}
        \item For every $p \in P$, and the facility $i'^* \in S'^*$ in $\ball_F(p, r_p)$, we have 
        \begin{align*}
            r_p \leq \frac{1+\epsilon}{1-\epsilon}\cdot\frac{\sum_{j \in \sigma^{-1}(i'^*)} d(i'^*, j)}{|\sigma^{-1}(i'^*)|}. 
        \end{align*} \label{property:r_p-to-sigma}
    \end{enumerate}
\end{lemma}
We remark that the balls $\ball_F(p, r_p)$ may overlap with each other and thus one facility in $P$ may be in two different balls. However, we guarantee that there is precisely one facility in $S'^*$ that is designated to $\ball_F(p, r_p)$ for each $p \in P$, and the facility is inside the ball. 

\begin{proof}[Proof of Lemma~\ref{lemma:desired-P-radius}]
    We construct the solution $S'^*$ with the assignment $\sigma$ as follows. For every optimum cluster $(i^*, J)$ with per-client-cost $\bar d$, we include a facility $i'^*$ in $S'^*$. Let $p$ be the desired pivot for this optimum cluster.  If $(i^*, J)$ is of type-1 or 2, we let $i'^* = i^*$. In this type, we have $i'^* = i^* \in \ball_F(p, r_p)$ as $r_p$ is the desired radius for $p$. 
    Otherwise, we are of type-3 and the pivot $p$ is the nearest facility in $S$ to $i^*$. We let $i'^* = p$ and thus $i'^* \in \ball(p, r_p = 0)$. In any type, we let $\sigma_j = i'^*$ for  every $j \in J$. \ref{property:r_p-to-sigma} holds if $(i^*, J)$ is of type-1 or 2 as $r_p \leq \frac{1+\epsilon}{1-\epsilon}\bar d$. of type-3, we have $r_p = 0$ and the property holds trivially. 

    It remains to show \ref{property:sigma-good}. 
    %We show $\cost_\topcn(\sigma)$ is small compare to $\cost_\topcn(S^*)$. 
    Notice that the difference between the left side of the inequality in \ref{property:sigma-good} and $\opt$ come from type-3 optimum clusters. Therefore, for every type-1 or 2 optimum cluster $(i^*, J)$, we define $\rho_j = 0$ for every $j \in J$. Fix a type-3 optimum cluster $(i^*, J)$, and its correspondent desired pivot $p$, we define $\rho_j = (d(j, p) - d(j, i^*))^+$ for every $j \in J$.
    \begin{align*}
        \sum_{j \in J} \rho_j &\leq |J|\cdot d(i^*, p) \leq |J|\cdot \frac{1}{|J'|} \left(\frac{\epsilon^3}{k} + \frac{\epsilon^2}{k}\right) |\vec{d}(S)|_1 \leq \frac{1}{\epsilon}\cdot \frac{2\epsilon^2}{k} |\vec{d}(S)|_1= \frac{2\epsilon}{k}|\vec{d}(S)|_1.
    \end{align*}
    The second inequality comes from triangle inequality and the definition of type-3. 
    
    Summing up over all type-3 optimum clusters, we have $|\rho|_1 \leq 2\epsilon|\vec{d}(S)|_1$. Therefore,
    \begin{align*}
        |\rho|_\topcn \leq |\rho|_1 \leq O(\epsilon)\cdot |\vec{d}(S)|_1 \leq O(\epsilon)\cdot |\vec{d}(S)|_{\topcn} \leq O(\epsilon)\cdot \opt.
    \end{align*}
    The third inequality used that $c \in  (\frac1e, 1]$. 
\end{proof}

\subsection{Step~\ref{step:Tpcn-return} of Algorithm~\ref{algo:top-cn}: Find Clustering using $P$ and $(r_p)_{p \in P}$}
In this final step, we try to find a clustering using the pivot set $P$ and the radius vector $(r_p)_{p \in P}$ as a guide, by calling $\texttt{TpcnC-clustering-with-pivots}(P, (r_p)_{p \in P})$ in Algorithm~\ref{alg:clustering-with-pivots}. The existence of a good clustering is guaranteed by Lemma~\ref{lemma:desired-P-radius}. 

%\texttt{clustering-with-pivots}

\begin{algorithm}[h]
    \caption{$\texttt{TpcnC-clustering-with-pivots}()$}
    \label{alg:clustering-with-pivots}
    \begin{algorithmic}[1]
        \State solve program \eqref{LP:with-pivots} \label{step:Tpcn-lp-pivots}
        to obtain solution $\big((x^{(p)}_{ij})_{p,i,j}, (y^{(p)}_i)_{p,i}, \delta \in \calS_n\big)$ \label{step:Tpcn-solve-lp-with-pivots}
        \For{every $p \in P$} randomly open a facility $i$, with probabilities $y^{(p)}_{i}$ \label{step:Tpcn-rounding}
        \EndFor
        \State \Return the set of open facilities 
    \end{algorithmic}
\end{algorithm}

% \begin{theorem}
%     If the algorithm is successful in Step xx and xx of Algorithm~\ref{algo:generic}, then it has an approximation ratio at most 
%     \begin{align*}
%         (1 + O(\epsilon)) \cdot \max_{\delta \in S_n}\frac{\overline{f}((1 - \frac1e)\delta + \frac1e \cdot (\delta \otimes 3))}{\overline{f}(\delta)}
%     \end{align*}
% \end{theorem}

%Then in Step~\ref{step:Tpcn-return}, we randomly open a facility from every $\ball_F(p, r_p), p \in P$, with probabilities defined by $y^{(p)}_i$ values. In Step~\ref{step:Tpcn-return}, we return the set of $k$ open facilities. In the actual algorithm, we return the best solution over all possible guesses.

In Step~\ref{step:Tpcn-lp-pivots} of Algorithm~\ref{alg:clustering-with-pivots}, we solve the following program:
\begin{align}
    %\min \qquad \sum_{p, i, j} (d(i, j) - t)^+ x^{(p)}_{ij} + cnt
    \min \qquad \overline{\topcn}(\delta) \label{LP:with-pivots}
\end{align}
\begin{align}
	\sum_{i \in \ball_F(p, r_p)} y^{(p)}_i &= 1 &\qquad &\forall p \in P \label{LP:open-1-facility}\\
	x^{(p)}_{ij}  &\leq y^{(p)}_i &\quad &\forall p \in P, i \in \ball_F(p, r_p), j \in C\label{LP:connect-to-open-ball}\\
	\sum_{p \in P, i \in \ball(p, r_p)} x^{(p)}_{ij} &= 1 &\quad &\forall j \in C\label{j-connect-open-i}\\
    \frac{1+\epsilon}{1-\epsilon}\sum_{j \in C} d(i, j) x^{(p)}_{ij} -  r_p \sum_{j \in C} x^{(p)}_{ij} &\geq 0 &\quad &\forall p \in P, i \in \ball_F(p, r_p)\label{LP:3.8b}\\
     \delta_a - \sum_{p, i,j:d(i,j) = a} x^{(p)}_{ij} &= 0  &\quad &\forall a \in [0, \Delta] \label{LPC:delta}\\
%     \delta'_a &= \sum_{p, i,j:d(i,p) = a} x^{(p)}_{ij},  &\quad &\forall a \in [0, \Delta];\\
	x^{(p)}_{ij}, y^{(p)}_i &\geq 0 &\quad &\forall p \in P, i \in F, j \in C\label{bicriteria-LP-end}
\end{align}

We now focus on the correspondent integer program to program (\ref{LP:with-pivots}), whose goal is to find the set $S'^*$ and the assignment $\sigma$ in Lemma~\ref{lemma:desired-P-radius}. $y^{(p)}_i$ indicates whether $i$ is the open facility in $\ball_F(p, r_p)$, and $x^{(p)}_{ij}$ indicates whether the client $j$ is connected to facility $i$ in $\ball_F(p, r_p)$. \eqref{LP:open-1-facility} indicates we open exactly one facility in $\ball_F(p, r_p)$, \eqref{LP:connect-to-open-ball} indicates that a client can only be connected to an open facility, and \eqref{j-connect-open-i} requires every client to be connected to a facility in all balls.  \eqref{LP:3.8b} is due to \ref{property:r_p-to-sigma}, and \eqref{LPC:delta} gives the definition of $\delta \in \calS_n$. 
%In the integer programming, we require $x^{(p)}_{ij}, y^{(p)}_i\in \{0, 1\}$ for every $p\in P, i\in F, j\in C$. In the LP(\ref{LP:with-pivots}) relaxation, we relax the constraint to $x^{(p)}_{ij}, y^{(p)}_i\in [0, 1]$.

We then describe the objective \eqref{LP:with-pivots}, which is not a linear function of $\delta$. By Lemma~\ref{lemma:linearize-top-l}, we know that $\overline{\topcn}(\delta) = \min_{t \in [0, \Delta]}\left(\sum_{a} (a - t)^+ \delta_{a} + cnt\right)$, we can solve the program by enumerating integers $t \in [0, \Delta]$. \footnote{By allowing a $1+\epsilon$ loss, we can assume $t$ is an integer power of $1+\epsilon$, and we only need to enumerate $O(\frac{\log n}{\epsilon})$ different values of $t$.} \smallskip

After solving the CP(\ref{LP:with-pivots}), we obtain the solution $\big((x^{(p)}_{ij})_{p,i,j}, (y^{(p)}_i)_{p,i}, \delta\in \calS_n\big)$.  Let $\lp'$ be the value of the solution. By Property~\ref{property:sigma-good}, we have $\lp' \leq (1 + O(\epsilon))\opt$. % that the value of the LP has a cost of at most $(1+O(\epsilon))\opt$.
%
 %Now we are going to use Theorem xx from Section xx. We need to define the occurrence-times vectors $\delta, \delta'$ and $\phi$:
% \begin{itemize}
%     \item Define $\delta \in \calS_n$ to be the occurrence vector for the LP solution: let $\delta_a = \sum_{p, i, j: d(i, j)=a} x^{(p)}_{ij}$ for every $a \in \Z_{\geq 0}$. That is, for every $p, i, j$, we include $x^{(p)}_{ij}$ fractional connection of distance $d(i, j)$ in $\delta$. By lemma XX, we have $\overline{\topcn}(\delta) \leq \lp'$. %TODO: state a lemma.
%     \item 
%     \item $z_{a, a'} := \sum_{p, i, j: d(i, j) = a, d(i, p) = a'} x^{(p)}_{ij}$ for every $a, a' \in \Z_{\geq 0}$. Let $\phi \in \calS_n$ be that $\phi_t = \sum_{a, a': a + 2a' = t } z_{a,a'} =  \sum_{p, i, j: d(i,j) + 2\floor{r_p} = t} x^{(p)}_{ij}$. So, in $\phi$, we include $x^{(p)}_{ij}$ factional connection of distance $d(i,j) + 2\floor{r_p}$ for every $p, i, j$. 
% \end{itemize}
%
Let $\psi$ be integral occurrence-times vector of the solution given by the algorithm. That is, for every $a \in \Z_{\geq 0}$, $\psi_a$ is the number of clients in $C$ with connection distance $a$. So our goal is to upper bound $\E[\overline{\topcn}(\psi)]$ in terms of $\lp'$. 

We define a vector $\delta' \in \calS_n$ as follows: $\delta'_{a'} = \sum_{p, i, j:\floor{r_p} = a'} x^{(p)}_{ij}$ for every $a' \in \Z_{\geq 0}$.  That is, for every $p, i, j$, we include $x^{(p)}_{ij}$ fractional connection of distance $\floor{r_p}$ in $\delta'$. By Corollary~\ref{coro:dominating-averaging} and \eqref{LP:3.8b}, we have 
\begin{align*}
    \delta' \preceq_{\frac{1+\epsilon}{1-\epsilon}} \delta.
\end{align*}
To see the inequality, we consider how $\delta$ and $\delta'$ is constructed: For every $p \in P$ and $i \in \ball_F(p, r_p)$, we include $x^{(p)}_{ij}$ fraction of distance $d(i, j)$ in $\delta$ for every $j$, and we include $\sum_{j \in C} x^{(p)}_{ij}$ fraction of distance $\floor{r_p} \leq \frac{1+\epsilon}{1-\epsilon}\sum_{j \in C}d(i, j)x^{(p)}_{ij}\big/ \sum_{j \in C}x^{(p)}_{ij}$ in $\delta'$. Thus Corollary~\ref{coro:dominating-averaging} can be applied.

The main lemma we prove is the following:
\begin{lemma}\label{lemma:bound-E-psi}
    $\E[\psi]$ is dominated by $\left(1-\frac1e\right)\delta + \frac1e\cdot(\delta\oplus\delta'\otimes 2)$.
\end{lemma}
We show why the lemma implies the desired approximation ratio. By the concavity of the $\overline{\topnorm{cn}}(\cdot)$ function in Lemma~\ref{lemma:concave}, we have 
\begin{align*}
    \E\left[\overline{\topnorm{cn}}(\psi)\right] &\leq \overline{\topnorm{cn}}(\E[\psi]) \leq \overline{\topnorm{cn}}\left(\big(1-\frac1e\big)\delta + \frac1e\cdot(\delta\oplus\delta'\otimes 2)\right) \\
    &\leq \left(1 + \frac{2}{ec} \cdot\frac{1+\epsilon}{1-\epsilon}\right)\cdot \overline{\topnorm{cn}(\delta)} \leq\left(1 + \frac{2}{ec} + O(\epsilon)\right)\opt.
\end{align*}

\begin{proof}[Proof of Lemma~\ref{lemma:bound-E-psi}]
    We focus on a single client $j \in C$. It is optimal to connect $j$ to the nearest open facility. But for the sake of analysis, it is convenient to connect $j$ to a random and possibly sub-optimal facility. Also, the connection cost we impose on $j$ could be larger than its actual cost.

    Abusing notations slightly, we use a pair $(p, i)$ to denote the copy of facility $i$ dedicated to $\ball_F(p, r_p)$. The probability that one facility in $\{(p,i): x^{(p)}_{ij} > 0\}$ is open is at least $1 - \frac1e$.  Using a contention resolution scheme,
    we can connect $j$ to 0 or 1 open facility in the set, such that 
    \begin{align*}
        \Pr[j\text { connected to }  (p, i)] = \left(1 - \frac1e\right)x^{(p)}_{ij}, \forall (p, i).
    \end{align*}
    When $j$ is connected to $(p, i)$, we impose a connection cost of $d(i ,j)$ for $j$. In this type, we say $j$ is \emph{directly} connected.

    With the remaining probability of $1 - \left(1-\frac1e\right)\sum_{i,p}x^{(p)}_{ij} = 1 - \left(1-\frac1e\right)=\frac1e$, we say $j$ is \emph{indirectly} connected. We then specify the connection cost in this type. For every facility $(p, i)$ with $x^{(p)}_{ij} > 0$, we know that some facility in $\ball_F(p, r_p)$ must be open. Then we can impose a connection cost of $d(i, j) + 2 \floor{r_p}$ on $j$ by connecting it to the facility in the ball. We make the indirect connection randomly using $x^{(p)}_{ij}$ values: conditioned on that we make an indirect connection for $j$, we randomly choose a pair $(p, i)$ with probabilities $x^{(p)}_{ij}$, and we impose a connection cost of $d(i, j) + 2 \floor{r_p}$ on $j$.

    Therefore, the connection cost of $j$ is distributed as follows:
    \begin{itemize}
        \item For every facility $(p, i)$, $j$ is directly connected to $(p, i)$ (and thus incurs a cost of $d(i,j)$) with probability $\left(1 - \frac1e\right)x^{(p)}_{ij}$.
        \item For every facility $(p, i)$, $j$ is indirectly connected via the facility $(p, i)$ (and thus incurs a cost of $d(i, j) + 2\floor{r_p}$) with probability $\frac1e\cdot x^{(p)}_{ij}$.
    \end{itemize}

    Now we consider all clients $j \in C$. We have that $\E[\psi]$ is dominated by $\big(1 - \frac1e\big)\delta + \frac1e\cdot \phi$ for some $\phi$ obtained by adding $\delta$ and $\delta'\otimes 2$, which is dominated by $\delta\oplus\delta'\otimes 2$ by Lemma~\ref{lemma:dominating-for-adding}. Therefore, by Lemma~\ref{lemma:dominating-adding-transitive}, we have $\E[\psi]$ is dominated by $\left(1-\frac1e\right)\delta + \frac1e\cdot(\delta\oplus\delta'\otimes 2)$.
\end{proof}

\subsection{Wrapping up}
Therefore, when all steps are successful, and all our guesses are correct, the algorithm returns a solution with expected $\topcn$ norm being at most $\left(1 + \frac{2}{ec} + O(\epsilon)\right)\opt$. 

There are $|S \cup R|^k$ different choices for $P$ in Step~\ref{step:Tpcn-P} of Algorithm~\ref{algo:top-cn}, and 
$O\left(\frac{\log n}{\epsilon}\right)^k$ different choices for $(r_p)_{p \in P}$ in Step~\ref{step:Tpcn-r}. The success probability of Steps~\ref{step:Tpcn-S} and \ref{step:Tpcn-R} in Algorithm~\ref{alg:choose-R-top-cn} are respectively $1 - \frac1{n^2}$ and $\frac{\epsilon^{3k}}{k^k}$. Therefore, if we enumerate all choices of $P$ and $(r_p)_{p \in P}$, and repeat the algorithm $O\left(\frac{k^k}{\epsilon^{3k}}\cdot \log n\right)$ times, the success probability can be increased to $1 - 1/\poly(n)$. Overall, we obtain a final algorithm with running time
\begin{align*}
    O\left(\frac{k\log n}{\epsilon^4}\right)^k \cdot \poly(n) \leq O\left(\frac{k^2\log k}{\epsilon^4}\right)^k \cdot \poly(n) \leq \poly\Big(k, \frac1\epsilon\Big)^k\cdot\poly(n),
\end{align*}
that with high probability outputs a solution whose $\topcn$ cost is at most $\left(1 + \frac{2}{ec} + O(\epsilon)\right)\cdot\opt$. To see the first inequality, notice that either $\log n \leq k \log k$, or $(\log n)^k \leq \poly(n)$; thus $(\log n)^k \leq \max\{(k \log k)^k, \poly(n)\}$.
    \bibliographystyle{plain}
    \bibliography{ICALP-article}

@InProceedings{DBLP:Cohenaddad2019,
  author =	{Cohen-Addad, Vincent and Gupta, Anupam and Kumar, Amit and Lee, Euiwoong and Li, Jason},
  title =	{{Tight FPT Approximations for $k$-Median and $k$-Means}},
  booktitle =	{46th International Colloquium on Automata, Languages, and Programming, {ICALP} 2019},
 volume =	{132},
  pages =	{42:1--42:14},
  year =	{2019}
}

@inproceedings{Alamdari2017,
  title={A Bicriteria Approximation Algorithm for the $k$-Center and $k$-Median Problems},
  author={Soroush Alamdari and David B. Shmoys},
  booktitle={Workshop on Approximation and Online Algorithms, WAOA 2017},
  volume       = {10787},
  pages        = {66--75},
  year={2017},
}

@article{Charikar2002,
  author       = {Moses Charikar and
                  Sudipto Guha and
                  {\'{E}}va Tardos and
                  David B. Shmoys},
  title        = {A Constant-Factor Approximation Algorithm for the $k$-Median Problem},
  journal      = {Journal of Computer and System Sciences},
  volume       = {65},
  number       = {1},
  pages        = {129--149},
  year         = {2002}
}

@inproceedings{Abbasi2023,
  author       = {Fateme Abbasi and
                  Sandip Banerjee and
                  Jaroslaw Byrka and
                  Parinya Chalermsook and
                  Ameet Gadekar and
                  Kamyar Khodamoradi and
                  D{\'{a}}niel Marx and
                  Roohani Sharma and
                  Joachim Spoerhase},
  title        = {Parameterized Approximation Schemes for Clustering with General Norm
                  Objectives},
  booktitle    = {64th {IEEE} Annual Symposium on Foundations of Computer Science, {FOCS}
                  2023},
  pages        = {1377--1399},
  year         = {2023}
}

@inproceedings{Chen2006,
  author       = {Ke Chen},
  title        = {On \emph{k}-Median clustering in high dimensions},
  booktitle    = {Proceedings of the 17th  Annual {ACM-SIAM} Symposium on Discrete
                  Algorithms, {SODA} 2006},
  pages        = {1177--1185},
  year         = {2006}
}

@article{Chen2009,
  author       = {Ke Chen},
  title        = {On Coresets for $k$-Median and $k$-Means Clustering in Metric and Euclidean Spaces and Their Applications},
  journal      = {SIAM Journal on Computing},
  volume       = {39},
  number       = {3},
  pages        = {923--947},
  year         = {2009}
}

@article{Feldman2020,
  author       = {Dan Feldman and
                  Melanie Schmidt and
                  Christian Sohler},
  title        = {Turning Big Data Into Tiny Data: Constant-Size Coresets for $k$-Means,
                  PCA, and Projective Clustering},
  journal      = {SIAM Journal on Computing},
  volume       = {49},
  number       = {3},
  pages        = {601--657},
  year         = {2020}
}

@inproceedings{Li2016,
  author       = {Shi Li},
  title        = {Approximating capacitated \emph{k}-median with $(1+\epsilon)k$ open facilities},
  booktitle    = {Proceedings of the 27th Annual {ACM-SIAM} Symposium on Discrete
                  Algorithms, {SODA} 2016},
  pages        = {786--796},
  year         = {2016},
}

@inproceedings{Adamczyk2019,
  author       = {Marek Adamczyk and
                  Jaroslaw Byrka and
                  Jan Marcinkowski and
                  Syed Mohammad Meesum and
                  Michal Wlodarczyk},
  title        = {Constant-Factor {FPT} Approximation for Capacitated $k$-Median},
  booktitle    = {27th Annual European Symposium on Algorithms, {ESA} 2019},
  volume       = {144},
  pages        = {1:1--1:14},
  year         = {2019}
}

@article{Goyal2023,
  author       = {Dishant Goyal and
                  Ragesh Jaiswal},
  title        = {Tight {FPT} Approximation for Socially Fair Clustering},
 journal={Information Processing Letters},
  volume={182},
  pages={106383},
  year={2023}
}

@inproceedings{Bandyapadhyay2023,
  author       = {Sayan Bandyapadhyay and
                  William Lochet and
                  Saket Saurabh},
  title        = {{FPT} Constant-Approximations for Capacitated Clustering to Minimize the Sum of Cluster Radii},
  booktitle = {39th International Symposium on Computational Geometry, SoCG 2023},
  volume       = {258},
  pages        = {12:1--12:14},
  year         = {2023}
}

@inproceedings{Feng2020,
  author       = {Qilong Feng and
                  Zhen Zhang and
                  Ziyun Huang and
                  Jinhui Xu and
                  Jianxin Wang},
  title        = {A Unified Framework of {FPT} Approximation Algorithms for Clustering
                  Problems},
  booktitle    = {31st International Symposium on Algorithms and Computation, {ISAAC}
                  2020},
  volume       = {181},
  pages        = {5:1--5:17},
  year         = {2020}
}

@article{Ahmadian2020,
  author       = {Sara Ahmadian and
                  Ashkan Norouzi{-}Fard and
                  Ola Svensson and
                  Justin Ward},
  title        = {Better Guarantees for $k$-Means and Euclidean $k$-Median by Primal-Dual
                  Algorithms},
  journal      = {SIAM Journal on Computing},
  volume       = {49},
  number       = {4},
  year         = {2020}
}

@article{Byrka2017, 
  author       = {Jaroslaw Byrka and
                  Thomas W. Pensyl and
                  Bartosz Rybicki and
                  Aravind Srinivasan and
                  Khoa Trinh},
  title        = {An Improved Approximation for \emph{k}-Median and Positive Correlation
                  in Budgeted Optimization},
  journal      = {ACM Transactions on Algorithms},
  volume       = {13},
  number       = {2},
  pages        = {23:1--23:31},
  year         = {2017}
}

@inproceedings{Charikar1999,
  author       = {Moses Charikar and
                  Sudipto Guha},
  title        = {Improved Combinatorial Algorithms for the Facility Location and $k$-Median Problems},
  booktitle    = {40th Annual Symposium on Foundations of Computer Science, {FOCS} 1999},
  pages        = {378--388},
  year         = {1999}
}

@inproceedings{Byrka2018,
  author       = {Jaroslaw Byrka and
                  Krzysztof Sornat and
                  Joachim Spoerhase},
  title        = {Constant-factor approximation for ordered $k$-median},
  booktitle    = {Proceedings of the 50th Annual {ACM} {SIGACT} Symposium on Theory
                  of Computing, {STOC} 2018},
  pages        = {620--631},
  year         = {2018}
}

@inproceedings{Chakrabarty2018,
  author       = {Deeparnab Chakrabarty and
                  Chaitanya Swamy},
  title        = {Interpolating between $k$-Median and $k$-Center: Approximation Algorithms
                  for Ordered $k$-Median},
  booktitle    = {45th International Colloquium on Automata, Languages, and Programming,
                  {ICALP} 2018},
  volume       = {107},
  pages        = {29:1--29:14},
  year         = {2018}
}

@inproceedings{Chakrabarty2019,
  author       = {Deeparnab Chakrabarty and
                  Chaitanya Swamy},
  title        = {Approximation algorithms for minimum norm and ordered optimization
                  problems},
  booktitle    = {Proceedings of the 51st Annual {ACM} {SIGACT} Symposium on Theory
                  of Computing, {STOC} 2019},
  pages        = {126--137},
  year         = {2019}
}

@article{Plesnik1987,
  author       = {J{\'{a}}n Plesn{\'{\i}}k},
  title        = {A heuristic for the $p$-center problems in graphs},
  journal      = {Discrete Applied Mathematics},
  volume       = {17},
  number       = {3},
  pages        = {263--268},
  year         = {1987}
}

@inproceedings{kanungo2002,
  title={A local search approximation algorithm for $k$-means clustering},
  author={Kanungo, Tapas and Mount, David M and Netanyahu, Nathan S and Piatko, Christine D and Silverman, Ruth and Wu, Angela Y},
  booktitle={Proceedings of the 18th Annual Symposium on Computational Geometry, SoCG 2002},
  pages={10--18},
  year={2002}
}

@article{Li2017uniform,
  title={On uniform capacitated $k$-median beyond the natural LP relaxation},
  author={Li, Shi},
  journal={ACM Transactions on Algorithms},
  volume={13},
  number={2},
  pages={1--18},
  year={2017}
}

@inproceedings{chuzhoy2005approximating,
  title={Approximating $k$-median with non-uniform capacities},
  author={Chuzhoy, Julia and Rabani, Yuval},
  booktitle={Proceedings of the 16th Annual ACM-SIAM Symposium on Discrete Algorithms, SODA 2005},
  volume={5},
  pages={952--958},
  year={2005}
}

@article{aardal2015approximation,
  title={Approximation algorithms for hard capacitated $k$-facility location problems},
  author={Aardal, Karen and L.van den Berg, Pieter  and Gijswijt, Dion and Li, Shanfei},
  journal={European Journal of Operational Research},
  volume={242},
  number={2},
  pages={358--368},
  year={2015}
}

@inproceedings{byrka2016approximation,
  title={An Approximation Algorithm for Uniform Capacitated $k$-Median Problem with $(1+\epsilon)$-Capacity Violation},
  author={Byrka, Jaros{\l}aw and Rybicki, Bartosz and Uniyal, Sumedha},
  booktitle={International Conference on Integer Programming and Combinatorial Optimization, {IPCO} 2016},
  volume = {9682},
  pages={262--274},
  year={2016}
}

@article{li2014improved,
  title={An improved approximation algorithm for the hard uniform capacitated $k$-median problem},
  author={Li, Shanfei},
  journal={arXiv preprint arXiv:1406.4454},
  year={2014}
}

@inproceedings{byrka2014bi,
  title={Bi-factor approximation algorithms for hard capacitated $k$-median problems},
  author={Byrka, Jaros{\l}aw and Fleszar, Krzysztof and Rybicki, Bartosz and Spoerhase, Joachim},
  booktitle={Proceedings of the 26h Annual ACM-SIAM Symposium on Discrete Algorithms, SODA 2015},
  pages={722--736},
  year={2014}
}

@inproceedings{demirci2016constant,
  title={Constant Approximation for Capacitated $k$-Median with $(1+ \epsilon)$-Capacity Violation},
  author={Demirci, G{\"o}kalp and Li, Shi},
  booktitle={43rd International Colloquium on Automata, Languages, and Programming, {ICALP} 2016},
volume       = {55},
  pages        = {73:1--73:14},
  year={2016}
}

@article{xu2019constant,
  title={A constant parameterized approximation for hard-capacitated $k$-means},
  author={Xu, Yicheng and M{\"o}hring, Rolf H and Xu, Dachuan and Zhang, Yong and Zou, Yifei},
  journal={arXiv preprint arXiv:1901.04628},
  year={2019}
}

@inproceedings{cygan2012lp,
  title={LP rounding for $k$-centers with non-uniform hard capacities},
  author={Cygan, Marek and Hajiaghayi, MohammadTaghi and Khuller, Samir},
  booktitle={53rd Annual IEEE Symposium on Foundations of Computer Science, FOCS 2012},
  pages={273--282},
  year={2012}
}

@article{khuller2000capacitated,
  title={The capacitated $k$-center problem},
  author={Khuller, Samir and Sussmann, Yoram J},
  journal={SIAM Journal on Discrete Mathematics},
  volume={13},
  number={3},
  pages={403--418},
  year={2000}
}

@article{an2015centrality,
  title={Centrality of trees for capacitated $k$-center},
  author={An, Hyung-Chan and Bhaskara, Aditya and Chekuri, Chandra and Gupta, Shalmoli and Madan, Vivek and Svensson, Ola},
  journal={Mathematical Programming},
  volume={154},
  number={1},
  pages={29--53},
  year={2015}
}

@article{barilan1993allocate,
  title={How to allocate network centers},
  author={Barilan, Judit and Kortsarz, Guy and Peleg, David},
  journal={Journal of Algorithms},
  volume={15},
  number={3},
  pages={385--415},
  year={1993}
}

@inproceedings{cohen2022fixed,
  title={On the Fixed-Parameter Tractability of Capacitated Clustering},
  author={Cohen-Addad, Vincent and Li, Jason},
  booktitle={46th International Colloquium on Automata, Languages, and Programming, ICALP 2019},
  volume={132},
  pages={41--1},
  year={2019}
}

@inproceedings{fakcharoenphol2003tight,
  title={A tight bound on approximating arbitrary metrics by tree metrics},
  author={Fakcharoenphol, Jittat and Rao, Satish and Talwar, Kunal},
  booktitle={Proceedings of the 35th Annual ACM Symposium on Theory of Computing, STOC 2003},
  pages={448--455},
  year={2003}
}

@inproceedings{arya2001local,
  title={Local search heuristic for $k$-median and facility location problems},
  author={Arya, Vijay and Garg, Naveen and Khandekar, Rohit and Meyerson, Adam and Munagala, Kamesh and Pandit, Vinayaka},
  booktitle={Proceedings of the 33rd Annual ACM Symposium on Theory of Computing, STOC 2001},
  pages={21--29},
  year={2001}
}

@article{byrka2017improved,
  title={An improved approximation for $k$-median and positive correlation in budgeted optimization},
  author={Byrka, Jaros{\l}aw and Pensyl, Thomas and Rybicki, Bartosz and Srinivasan, Aravind and Trinh, Khoa},
  journal={ACM Transactions on Algorithms},
  volume={13},
  number={2},
  pages={1--31},
  year={2017}
}

@article{jain2001approximation,
  title={Approximation algorithms for metric facility location and $k$-median problems using the primal-dual schema and Lagrangian relaxation},
  author={Jain, Kamal and Vazirani, Vijay V},
  journal={Journal of the ACM},
  volume={48},
  number={2},
  pages={274--296},
  year={2001}
}

@inproceedings{cohen2022improved,
  title={An improved local search algorithm for $k$-median},
  author={Cohen-Addad, Vincent and Gupta, Anupam and Hu, Lunjia and Oh, Hoon and Saulpic, David},
  booktitle={Proceedings of the 2022 Annual ACM-SIAM Symposium on Discrete Algorithms, SODA 2022},
  pages={1556--1612},
  year={2022}
}

@article{gupta2008simpler,
  title={Simpler analyses of local search algorithms for facility location},
  author={Gupta, Anupam and Tangwongsan, Kanat},
  journal={arXiv preprint arXiv:0809.2554},
  year={2008}
}

@article{lloyd1982least,
  title={Least squares quantization in PCM},
  author={Lloyd, Stuart},
  journal={IEEE Transactions on Information Theory},
  volume={28},
  number={2},
  pages={129--137},
  year={1982}
}

@inproceedings{cohen20252+,
  title={A $(2+ \epsilon)$-Approximation Algorithm for Metric $k$-Median},
  author={Cohen-Addad, Vincent and Grandoni, Fabrizio and Lee, Euiwoong and Schwiegelshohn, Chris and Svensson, Ola},
  booktitle={Proceedings of the 57th Annual ACM Symposium on Theory of Computing, STOC 2025},
  pages={615--624},
  year={2025}
}

@article{osadnik2023fixed,
  title={Fixed Parameter Tractable Algorithm and Coreset for the Ordered $k$-Median problem, Master thesis},
  author={Osadnik, Michal},
  year={2023},
  journal={https://urn.fi/URN:NBN:fi:aalto-202405193486}
}

@article{hochbaum1986unified,
  title={A unified approach to approximation algorithms for bottleneck problems},
  author={Hochbaum, Dorit S and Shmoys, David B},
  journal={Journal of the ACM},
  volume={33},
  number={3},
  pages={533--550},
  year={1986}
}

@article{guha1999greedy,
  title={Greedy strikes back: Improved facility location algorithms},
  author={Guha, Sudipto and Khuller, Samir},
  journal={Journal of Algorithms},
  volume={31},
  number={1},
  pages={228--248},
  year={1999}
}

@article{hochbaum1985best,
  title={A best possible heuristic for the $k$-center problem},
  author={Hochbaum, Dorit S and Shmoys, David B},
  journal={Mathematics of Operations Research},
  volume={10},
  number={2},
  pages={180--184},
  year={1985}
}

@article{gonzalez1985clustering,
  title={Clustering to minimize the maximum intercluster distance},
  author={Gonzalez, Teofilo F},
  journal={Theoretical Computer Science},
  volume={38},
  pages={293--306},
  year={1985}
}

@article{cohen2021near,
  title={Near-linear time approximation schemes for clustering in doubling metrics},
  author={Cohen-Addad, Vincent and Feldmann, Andreas Emil and Saulpic, David},
  journal={Journal of the ACM},
  volume={68},
  number={6},
  pages={1--34},
  year={2021}
}

@inproceedings{feder1988optimal,
  title={Optimal algorithms for approximate clustering},
  author={Feder, Tom{\'a}s and Greene, Daniel},
  booktitle={Proceedings of the 21th Annual ACM Symposium on Theory of Computing,  STOC 1988},
  pages={434--444},
  year={1988}
}

@inproceedings{eisenstat2014approximating,
  author       = {David Eisenstat and
                  Philip N. Klein and
                  Claire Mathieu},
  title        = {Approximating \emph{k}-center in planar graphs},
  booktitle    = {Proceedings of the 25th Annual {ACM-SIAM} Symposium on Discrete
                  Algorithms, SODA 2014},
  pages        = {617--627},
  year={2014}
}

@inproceedings{conf/focs/JainV99,
  author       = {Kamal Jain and
                  Vijay V. Vazirani},
  title        = {Primal-Dual Approximation Algorithms for Metric Facility Location
                  and $k$-Median Problems},
  booktitle    = {40th Annual Symposium on Foundations of Computer Science, {FOCS} 1999},
  pages        = {2--13},
  year         = {1999}
}

@inproceedings{cohen2022improvededuclidean,
  title={Improved approximations for euclidean $k$-means and $k$-median, via nested quasi-independent sets},
  author={Cohen-Addad, Vincent and Esfandiari, Hossein and Mirrokni, Vahab and Narayanan, Shyam},
  booktitle={Proceedings of the 54th Annual ACM SIGACT Symposium on Theory of Computing, {STOC} 2022},
  pages={1621--1628},
  year={2022}
}

@inproceedings{feldman2007ptas,
  title={A PTAS for $k$-means clustering based on weak coresets},
  author={Feldman, Dan and Monemizadeh, Morteza and Sohler, Christian},
  booktitle={Proceedings of the 23rd Annual Symposium on Computational Geometry, SoCG 2007},
  pages={11--18},
  year={2007}
}

@article{friggstad2019local,
  title={Local search yields a PTAS for $k$-means in doubling metrics},
  author={Friggstad, Zachary and Rezapour, Mohsen and Salavatipour, Mohammad R},
  journal={SIAM Journal on Computing},
  volume={48},
  number={2},
  pages={452--480},
  year={2019}
}

@inproceedings{lattanzi2019better,
  title={A better $k$-means++ algorithm via local search},
  author={Lattanzi, Silvio and Sohler, Christian},
  booktitle    = {Proceedings of the 36th International Conference on Machine Learning,
                  ICML 2019},
  volume       = {97},
  pages        = {3662--3671},
  year={2019}
}

@article{tcs/GoyalJ23,
  author       = {Dishant Goyal and
                  Ragesh Jaiswal},
  title        = {Tight {FPT} approximation for constrained \emph{k}-center and \emph{k}-supplier},
  journal      = {Theoretical Computer Science},
  volume       = {940},
  pages        = {190--208},
  year         = {2023}
}

@inproceedings{conf/icml/BravermanJKW19,
  author       = {Vladimir Braverman and
                  Shaofeng H.{-}C. Jiang and
                  Robert Krauthgamer and
                  Xuan Wu},
  title        = {Coresets for Ordered Weighted Clustering},
  booktitle    = {Proceedings of the 36th International Conference on Machine Learning,
                  {ICML} 2019},
  volume       = {97},
  pages        = {744--753},
  year         = {2019}
}

@inproceedings{conf/focs/BravermanCJKST022,
  author       = {Vladimir Braverman and
                  Vincent Cohen{-}Addad and
                  Shaofeng H.{-}C. Jiang and
                  Robert Krauthgamer and
                  Chris Schwiegelshohn and
                  Mads Bech Toftrup and
                  Xuan Wu},
  title        = {The Power of Uniform Sampling for Coresets},
  booktitle    = {63rd {IEEE} Annual Symposium on Foundations of Computer Science, {FOCS} 2022,},
  pages        = {462--473},
  year         = {2022}
}

@inproceedings{conf/soda/Huang0L025,
  author       = {Lingxiao Huang and
                  Jian Li and
                  Pinyan Lu and
                  Xuan Wu},
  title        = {Coresets for Constrained Clustering: General Assignment Constraints
                  and Improved Size Bounds},
  booktitle    = {Proceedings of the 2025 Annual {ACM-SIAM} Symposium on Discrete Algorithms,
                  {SODA} 2025},
  pages        = {4732--4782},
  year         = {2025}
}

@article{liu2025fixed,
  title={A Fixed-Parameter Tractable Approximation for Capacitated $k$-Supplier},
  author={Liu, Shuilian and Chen, Xianrun and Xu, Yicheng},
  journal={Theoretical Computer Science},
  pages={115605},
  year={2025}
}

@inproceedings{DBLP:CharikarCGGLW25,
  author       = {Moses Charikar and
                  Vincent Cohen{-}Addad and
                  Ruiquan Gao and
                  Fabrizio Grandoni and
                  Euiwoong Lee and
                  Ernest van Wijland},
  title        = {An Improved Greedy Approximation for (Metric) $k$-Means},
  booktitle    = {66th {IEEE} Annual Symposium on Foundations of Computer Science, {FOCS}
                   2025},
  pages        = {233--240},
  year         = {2025}
}

@article{DBLP:journals/dam/Tamir01,
  author       = {Arie Tamir},
  title        = {The $k$-centrum multi-facility location problem},
  journal      = {Discret. Appl. Math.},
  volume       = {109},
  number       = {3},
  pages        = {293--307},
  year         = {2001}
}

@article{DBLP:journals/mp/AouadS19,
  author       = {Ali Aouad and
                  Danny Segev},
  title        = {The ordered $k$-median problem: surrogate models and approximation algorithms},
  journal      = {Math. Program.},
  volume       = {177},
  number       = {1-2},
  pages        = {55--83},
  year         = {2019}
}
    \appendix
     \section{Making Distances Polynomially Bounded}
\label{appendix:distance-polynomial}

In this section, we show how to make distances $\infty$ or integers in $[0, \poly(n)]$, losing only a $(1+\epsilon)$-factor in the approximation ratio, for any symmetric monotone objective norm $f$. We make a guess on the maximum connection distance $L$ in the optimum solution. Consider the complete bipartite graph between $F$ and $C$ where the lengths of edges are the distances. If an edge has cost more than $L$, we remove the edge. Otherwise, we round the length down to the nearest integer multiple of $\frac{\epsilon L}{3n^2}$. The new distance is defined as the metric-completion of the graph. If the original distance between two vertices is at most $L$, then the new distance is in $[L - \frac{2\epsilon L}{3n}, L]$. The optimum cost of the two instances differ by at most the $f(\frac{2\epsilon L}{3n}, \frac{2\epsilon L}{3n}, \cdots, \frac{2\epsilon L}{3n})$, which is at most $\frac{2\epsilon}3 f(L, 0, 0, \cdots, 0) \leq \frac{2\epsilon}{3}\cdot \opt$. After scaling the distances by $\frac{3n^2}{2\epsilon L}$, all distances are $\infty$ or integers in $[0, \frac{n^3}{\epsilon}]$.

\section{Proof of Theorem~\ref{thm:pseudo-approx}} \label{appendix:pseudo-approx}

In this section, we prove Theorem~\ref{thm:pseudo-approx} and Corollary~\ref{coro:three-approximation}, which are repeated below for convenience. 
\MainThm*

We solve convex program CP\eqref{CP:f-logn},  defined as follows. 
\begin{align}
    \min \qquad  h(d^\av) \label{CP:f-logn}
\end{align}\vspace*{-30pt}

\noindent\begin{minipage}[t]{0.43\textwidth}
    \begin{align}
        \sum_{i \in F} y_i &= k \label{LPC:open-k}\\
        x_{ij}  &\leq y_i &\quad &\forall i \in F, j \in C \label{LPC:connect-to-open}\\
        \sum_{i \in F} x_{ij} &= 1 &\quad &\forall j \in C \label{LPC:j-connected}\\
        \sum_{j \in C} x_{ij} &\leq u_iy_i &\quad &\forall i \in F \label{LPC:capacity}
    \end{align}
\end{minipage}\hfill
\begin{minipage}[t]{0.52\textwidth}
    \begin{align}
        d^\av_j &= \sum_{i \in F}d(i, j)x_{ij} &\quad &\forall j \in C \label{LPC:define-dav}\\
        x_{ij} &\geq 0 &\quad &\forall i \in F, j \in C \label{LPC:x-non-negative}\\
        y_i &\geq 0 &\quad &\forall i \in F \label{LPC:y-non-negative}
    \end{align}
\end{minipage}\bigskip

         In integer program correspondent to CP(\ref{CP:f-logn}), $y_i$ indicates whether facility $i$ is open or not, and $x_{ij}$ indicates whether the client $j$ is connected to facility $i$. \eqref{LPC:open-k} restricts us to open $k$ facilities, \eqref{LPC:connect-to-open} requires that a client can only be connected to an open facility, \eqref{LPC:j-connected} requires every client to be connected to a facility, \eqref{LPC:capacity} is the capacity constraint, \eqref{LPC:define-dav} defines the variables $d^\av_j$'s, and \eqref{LPC:x-non-negative} and \eqref{LPC:y-non-negative} are the non-negativity constraints. 
         
         In the integer programming, we require $x_{ij}, y_i\in \{0, 1\}$ for every $i\in F, j\in C$. $d^\av_j$ is just the connection distance of $j$. 
         In CP(\ref{CP:f-logn}) relaxation; we relax the constraint to $x_{ij}, y_i\in [0, 1]$.  Then $d^\av_j$ becomes the average connection distance of $j$ over all its fractional connections.  The objective is to minimize $h(d^\av)$, which is convex. So the program is a convex program.

We solve CP~\eqref{CP:f-logn} to obtain a solution $(x, y, d^\av)$; let $\lp$ be the value of the solution. The algorithm is given in Algorithm~\ref{alg:pseudo}.

\bgroup
\algtext*{Until}{}		
\begin{algorithm}[h]
    \caption{$\texttt{Pseudo-Approximation}()$} \label{alg:pseudo}
    \begin{algorithmic}[1]
        \State solve CP\eqref{CP:f-logn} %(or a variant if $f = \ell_\infty$) 
        to obtain $\big((x_{ij})_{i\in F, j \in C}, (y_i)_{i \in F}, (d^\av_j)_{j \in C}\big)$ \label{step:solve-CP}
        \State $\calS \gets \emptyset$ \label{step:construct-S-start}
        \Repeat {$\frac{c'k\log n}{\epsilon}$ %(or $c'k\log n$ if $f \equiv \ell_\infty$) 
        times for a sufficiently large constant $c' > 0$}
            \State randomly choose a facility $i \in F$ according to the distribution $\{\frac{y_i}{k}\}_{i \in F}$
            \State choose a set $J \subseteq C$ randomly such that
                \begin{itemize}
                    \item $|J| \leq u_i$ with probability 1
                    \item $\Pr[j \in J] = \frac{x_{ij}}{y_i}$ for every $j \in C$
                \end{itemize}\Comment{It is well-known that such distribution $J$ exists and we can efficiently generate $J$ according to the distribution. }
            \State add $(i, J)$ to $\calS$
        \Until{}
        \State if a client $j$ appears in multiple stars in $\calS$, we only keep it in the star $(i, J)$ with the smallest $d(i, j)$
        \State \Return $\calS$
    \end{algorithmic}
\end{algorithm}
\egroup

\begin{observation}\label{obs:new-sol-opt}
    Each $(i, J) \in \calS$ is a valid star. 
\end{observation}

\begin{definition}\label{def: S is good}
    We say a client $j$ is good if there is a star $(i, J) \in \calS$ with $j \in J$ such that $d(i, j) \leq \frac{d^\av_j}{1-\epsilon}$.
\end{definition}

\begin{lemma}\label{lem: prob of S is good}
    With probability at least $1 - \frac1{n^2}$, all clients in $C$ are good.
\end{lemma}
\begin{proof}
    We say $j$ is covered in an iteration of the repeat loop, if in the iteration, we sampled a star $(i, J)$ with $j \in J$ and $d(i, j) \leq \frac{d^\av_j}{1-\epsilon}$. Notice that by Markov inequality and that $\sum_{i \in F} y_i = k$, the probability that $j$ is covered in an iteration is at least $\frac{\epsilon}{k}$. The probability that a client $j$ is not covered in any iteration is at most 
    \begin{align*}
        \left(1 - \frac{\epsilon}{k}\right)^{\frac{c'k\log n}{\epsilon}} \leq e^{-c' \log n} \leq \frac{1}{n^3},
    \end{align*}
    when $c'$ is sufficiently large. The lemma follows from the union bound over all clients $j \in C$. 
\end{proof}

Theorem~\ref{thm:pseudo-approx} follows from that $h$ is monotone. 

%For every $j \in C$, let $b_j = \min_{(i, J) \in \calS: j \in J} d(i, j)$. Then, 
% \begin{lemma}label{lem:(1+epsilon)-approx}
%     If all clients are good, then $h(b) \leq \frac{\lp}{1-\epsilon}$. % and $\cost_f(S) \leq \frac{L}{1-\epsilon}$.
% \end{lemma}
% \begin{proof}
%     When all clients are good, we have $b_j \leq \frac{d^\av_j}{1-\epsilon}$ for every $j \in C$. we have $h(b) \leq \frac{h(d^\av)}{1-\epsilon} = \frac{\lp}{1-\epsilon}$.
% \end{proof}

%Now, we show how to obtain the improved parameters when $f \equiv \ell_\infty$. In the convex program, we do not have an objective; thus we do not need the definition of $d^\av_j$'s. Instead, we guess the value $\opt$, and require $x_{ij} = 0$ whenever $d(i, j) > \opt$. (This is the variant of the convex program mentioned in Algorithm~\ref{alg:pseudo}.) In the algorithm, we only repeat the repeat loop $c'k\log n$ iterations.  We say a client $j$ is good whenever we sampled a star $(i, J)$ with $j \in J$. Now, $j$ is covered in an iteration with probability $\frac1k$. Therefore, $c'k\log n$ iterations suffice. When $\calS$ is good, the returned solution has $\ell_\infty$ norm at most $\opt$. 

\medskip
Then we prove Corollary~\ref{coro:three-approximation}, which is repeated below:
\corothreeapprox*
\begin{proof}
    We let $h$ be the norm $f$, and the capacities of facilities be $\infty$. We use Theorem~\ref{thm:pseudo-approx} to obtain the set $\calS$ of valid stars. Let $S$ be the facilities used by $\calS$. Then we enumerate all sets $T \subseteq S$ of size $k$ and output the best one. 

    We then analyze the approximation ratio of the algorithm. Let $S^*$ be the optimum set of $k$ facilities. For every $i^* \in S^*$, we let $\phi(i^*)$ be the facility in $S$ that is nearest to $i^*$.  We claim that $T:=\{\phi(i^*): i^* \in S^*\}$ gives a $(3+\epsilon)$ approximation for the problem. For every $j \in C$, let $i^*_j$ be the facility $j$ is connected to in the solution $S^*$. Let $i_j^S$ be the facility in $S $ closest to $j$.  Then we have 
    \begin{align*}
        d(j, T) &\leq d(j, i^*_j) + d(i^*_j, T) \leq d(j, i^*_j) + d(i^*_j, \phi(i^*_j)) \leq d(j, i^*_j) + d(i^*_j, j) + d(j, i_j^S)\\
        &\leq 2d(j, i^*_j) + d(j, S).
    \end{align*}
    The  third inequality holds by $d(i_j^*, \phi(i_j^*))\leq d(i_j^*, i_j^S)$ and the triangle inequality.
    Therefore, the connection distance of $j$ in the solution $T$ is at most twice its connection distance in the optimum solution, plus the connection distance in the solution $S$. So, its $f$-norm cost is at most $2\cdot\opt + \frac{\opt}{1-\epsilon} = (3+O(\epsilon))\opt$. 
\end{proof}

\section{Minimum-Norm Capacitated $k$-Clustering With Given Open Facilities} \label{appendix:find-assignment}

This section is devoted to the proof of Theorem~\ref{thm:find-assignment}, which is repeated below for convenience:
\thmfindassignment*

%In this section, we provide a $(1+\epsilon)$-approximation algorithm for capacitated $k$-clustering problem with arbitrary symmetric monotone norm, when the $k$ facilities are already given (i.e. $|F| = k$), for any $0 < \epsilon < \frac{1}{10}$. 
% Notice that, for many special problems, such as $k$-center and $k$-median, it is trivial to find the optimal clusters given the facilities.

\subsection{Preliminary}

We first provide important definitions useful for this section. Fix $0 < \epsilon < \frac{1}{10}$. For $c \in C$ and $\emptyset \subsetneq X \subseteq F$, say $c$ is \emph{$X$-exclusive}, if the following two conditions hold: \begin{itemize}
    \item[(i)] $\forall f_1, f_2 \in X, \frac{d(f_1,f_2)}{d(c, X)} \le \epsilon$;
    \item[(ii)] $\forall f\in F \backslash X, \frac{\max_{f' \in X} d(c, f')}{d(c, f)} \le \epsilon$.
\end{itemize}

If $c \in C$ is not $X$-exclusive for any $\emptyset \subsetneq X \subseteq F$, say $c$ is \emph{inclusive}. Define $C_X$ to be the set of $X$-exclusive clients, and $C_I$ as the set of inclusive clients.

We first show that, those $X$ with nonempty $C_X$ forms a laminar family, whose inclusion relation forms a tree structure. 

\begin{lemma}\label{lem:laminar}
    $\{X \mid C_X \neq \emptyset\}$ is a laminar family, i.e. If $X \cap Y \ne \emptyset$, then either $X \subseteq Y$ or $Y \subseteq X$.
\end{lemma}
\begin{proof}
    Suppose the contrary, assuming there exists $X$ and $Y$ satisfying $C_X \ne \emptyset$, $C_Y \ne \emptyset$, $X \cap Y \ne \emptyset$, $X \backslash Y \ne \emptyset$ and $Y \backslash X \ne \emptyset$. Pick $c_X \in C_X$, $c_Y \in C_Y$, $f_X \in X \backslash Y$, $f_Y \in Y \backslash X$ and $f_\cap \in X \cap Y$.

    Applied to set $X$, (i) shows $d(c_X,f_\cap) \ge d(c_X,X) \ge \frac{1}{\epsilon} d(f_X,f_\cap)$, and (ii) further shows $d(c_X,f_Y) \ge \frac{1}{\epsilon} \max_{f'\in X} d(c,f') \ge \frac{1}{\epsilon} d(c_X,f_\cap).$ By triangular inequality, $d(f_Y,f_\cap) \ge d(f_Y,c_X) - d(c_X, f_\cap) \ge \frac{1-\epsilon}{\epsilon} d(c_X,f_\cap) \ge \frac{1-\epsilon}{\epsilon^2} d(f_X,f_\cap)$.

    Similarly, $d(f_X,f_\cap) \ge \frac{1-\epsilon}{\epsilon^2}d(f_Y,f_\cap)$, a contradiction as $ \frac{1-\epsilon}{\epsilon^2} > 1$ for $\epsilon < \frac{1}{10}$.
\end{proof}
\begin{corollary}\label{lem:laminar-size}
    $|\{X \mid C_X \neq \emptyset\}| = O(k).$
\end{corollary}

% Now we show several properties regarding the definition. 

% The two lemmas are straightforward from triangular inequality.
% \begin{lemma}
%     If $c$ is $X$-exclusive, $\forall f_1, f_2 \in X$, $(1-\epsilon)d(c,f_2) \le d(c,f_1) \le (1+\epsilon) d(c,f_2)$
% \end{lemma}
% \begin{lemma}
%     If $c$ is $X$-exclusive, then $\forall f \in F \backslash X, (1-\epsilon) d(f,X) \le d(c,f) \le (1+\epsilon) d(f,X)$.
% \end{lemma}

In the following lemma, we show that $\{C_X \mid \emptyset \subsetneq X \subseteq F\}$ and $C_I$ partitions $C$.

\begin{lemma}\label{lem:exclusive-disjoint}
    For $ \emptyset \subsetneq X, Y \subseteq F$, $C_X \cap C_Y \ne \emptyset$ implies $X = Y$.
\end{lemma}
\begin{proof}
    Prove by contradiction. Choose any $c \in C_X \cap C_Y$. Without loss of generality, suppose $X \backslash Y \ne \emptyset$ and choose any $f_1 \in X \backslash Y$ and $f_2 \in Y$. (ii) applied to $Y$ gives $\frac{d(c,f_2)}{d(c,f_1)} \le \epsilon$. If $f_2 \in Y \backslash X$, (ii) applied to $X$ gives $\frac{d(c,f_1)}{d(c,f_2)}\le\epsilon$, a contradiction. Otherwise, $f_2 \in X$, (i) applied to $X$ gives $\frac{d(f_1,f_2)}{d(c,f_1)}\le \epsilon$. According to triangular inequality, $\frac{d(c,f_2)}{d(c,f_1)} \ge 1-\epsilon$, a contradiction with $\epsilon < \frac{1}{10}$.
\end{proof}

Finally, we show that, if a client is inclusive, then its distance to any facility is close to the distance between this facility and some other facility. Then, after discretization, there are only $O(k)$ possible distances for each facility.

\begin{lemma}\label{lem:invariant-necessary}
 If $c \in C_I$, then for every $ f_1 \in F$, there exists $f_2 \in F$, such that $\frac{\epsilon}{4} \le \frac{d(c,f_1)}{d(f_1,f_2)} \le \frac{4}{\epsilon}$.
\end{lemma}
\begin{proof}
    Suppose the contrary. Then there exists $f_1\in F$ such that, for every $f_2\in F$, $\frac{d(f_1, f_2)}{d(c, f_1)}<\frac{\epsilon}{4}$ or $\frac{d(f_1, f_2)}{d(c, f_1)}>\frac{4}{\epsilon}$.  Let $X: =\{f_2\in F: \frac{d(f_1, f_2)}{d(c, f_1)}<\frac{\epsilon}{4}\}$. We show that $c$ is $X$-exclusive, contradicting $c\in C_I$.  First, for every $x, y\in X$, $d(x, y)\leq d(x, f_1)+d(f_1, y)\leq \frac{\epsilon}{2}d(c, f_1)$. Then $d(c, X)=\min_{x\in X}d(c, x)\geq (1-\frac{\epsilon}{4})d(c, f_1)$, the inequality holds by $d(f_1, x)<\frac{\epsilon}{4} d(c, f_1)$ for every $x\in X$ and the triangle inequality. Thus condition (i) holds.  Second, for every $f\in F\setminus X$, we have $d(f_1, f)> \frac{4}{\epsilon}d(c, f_1)$. For every $x\in X$, $d(c, x)\leq (1+\frac\epsilon{4})d(c, f_1)$ and $d(c, f_1)\geq (\frac{4}{\epsilon}-1)d(c, f_1)$. Thus condition (ii) also holds, so $c\in C_X$.
\end{proof}

\begin{corollary}\label{cor:discretization}
    For any $f \in F$, $|\{(1+\epsilon)^{\ceil{\log_{1+\epsilon} d(c,f)}} \mid c \in C_I\}| = O(k)$.
\end{corollary}

\begin{proof}
    Lemma \ref{lem:invariant-necessary} says that, for all $f \in F$, $c \in C_I$, $d(c,f)\in \bigcup_{f' \in F} \left[\frac{\epsilon}{4} d(f,f'),\frac{4}{\epsilon} d(f,f')\right]$. For each interval, the discretization has $\log_{1+\epsilon}\frac{16}{\epsilon^2} = \frac{1}{\epsilon^{O(1)}} = O(1)$ possible results. Summing over all $f'\in F$, we conclude $|\{(1+\epsilon)^{\ceil{\log_{1+\epsilon} d(c,f)}} \mid c \in C_I\}|  = \frac{k}{\epsilon^{O(1)}}$. In particular, for fixed $\epsilon$, this number is $O(k)$.
\end{proof}

\subsection{Algorithm}

We first provide the intuition of the algorithm. Suffer $\epsilon$ loss to apply the discretization between distances from $C_I$ to $F$. Now, consider inclusive and exclusive clients separately. 
\begin{itemize}
    \item As Corollary \ref{cor:discretization} shows, after discretization, each facility has $O(k)$ possible distances to $C_I$. As the norm is symmetric, only the number of linkages for each distance matters, so in total $O(k^2)$ integers with value range $[0,n]$ suffice to determine their contribution.
    \item For $X$-exclusive clients, (ii) implies that linkages to any facility $f \in F \backslash X$ have distance roughly $d(f,X)$, with $\epsilon$ error. Therefore, only the number of linkages matters. 
    \item (i) further implies that, any $c \in C_X$ that links to $X$ has distance approximately $d(c,X)$, with $\epsilon$ loss. Since $F \backslash X$ pose no except frequency constraint on the matching from $C_X$, to minimize the cost from $C_X$ to $X$, we can always match the nearest clients in $C_X$ to $X$. Finally, only the number of linkages inside $X$ is useful.
\end{itemize}

The following lemma formalizes the intuition above.

\begin{lemma}\label{lem:info}
    For solutions $\mathcal{S}$ and $\mathcal{S}'$, $\mathrm{val}(\mathcal{S}') \le (1+O(\epsilon))\mathrm{val}(\mathcal{S})$ if $\mathcal{S}'$ satisfies all conditions below: \begin{itemize}
        \item $\forall f \in F, d \in \{(1+\epsilon)^{\lceil \log_{1+\epsilon}d(c,f) \rceil} \mid c \in C_I\},$ the number of linkages from $C_I$ to $f$ with discretized distance $d$ is the same between $\mathcal{S}$ and $\mathcal{S}'$;
        \item $\forall \emptyset \subsetneq X \subseteq F, \exists d_0 \in \mathbb{R}$, such that $\forall c \in C_X$, $c$ links to $X$ in $\mathcal{S}'$ if and only if $d(c,X) \le d_0$. \footnote{Without loss of generality, suppose $d(c,X)$ are distinct for $c \in C_X$.}
        \item $\forall \emptyset \subsetneq X \subseteq F, f \in F \backslash X$, the number of linkages from $C_X$ to $f$ is the same between $\mathcal{S}$ and $\mathcal{S}'$;
    \end{itemize}
\end{lemma}
\begin{proof}
    Those conditions imply that, we can always find a bijection $\phi: C \to C$, where for all $c \in C$, suppose $c$ matches $f$ in $\mathcal{S}$ and $\phi(c)$ matches $f'$ in $\mathcal{S}'$, then \begin{itemize}
        \item If $c \in C_I$, then $\phi(c) \in C_I$, $f' = f$ and $(1+\epsilon)^{\lceil \log_{1+\epsilon} d(c,f) \rceil} = (1+\epsilon)^{\lceil \log_{1+\epsilon} d(\phi(c),f') \rceil}$;
        \item If $c \in C_X$ and $f \in X$, then $\phi(c) \in C_X$, $f' \in X$ and $d(\phi(c),X) \le d(c,X)$;
        \item If $c \in C_X$ and $f \not\in X$, then $\phi(c) \in C_X$ and $f' = f$.
    \end{itemize}
    To prove the statement of the lemma, it suffices to show $d(\phi(c), f') \le (1+O(\epsilon))d(c,f)$. For the case $c \in C_I$, $(1+\epsilon)^{\lceil \log_{1+\epsilon} d(c,f) \rceil} = (1+\epsilon)^{\lceil \log_{1+\epsilon} d(\phi(c),f') \rceil}$ directly implies $d(\phi(c),f') \le (1+\epsilon) d(c,f)$. For the case $c \in C_X$ and $f \not\in X$, we have $f'= f$. By condition (ii), $d(\phi(c), f') \le \frac{1}{1-\epsilon} d(f', X) = \frac{1}{1-\epsilon} d(f, X) \le \frac{1+\epsilon}{1-\epsilon} d(c, f)$. For the case $c \in C_X$ and $f \in X$, condition (i) implies $d(\phi(c),f') \le (1+\epsilon) d(\phi(c),X) \le (1+\epsilon) d(c,X) \le (1+\epsilon) d(c,f)$. 
\end{proof}

Lemma \ref{lem:info} shows that, to construct an $(1+O(\epsilon))$-approximate solution, it suffices to know at most $O(k^2)$ integers in $[0,n]$ about the optimal solution. So the algorithm first guesses such information, and then constructs the solution based on the guess. The algorithm first guesses \begin{itemize}
    \item $\forall f \in F, d \in \{(1+\epsilon)^{\lceil \log_{1+\epsilon} d(c,f) \rceil} \mid c \in C_I\}$, the number of clients in $C_I$ matched to $f$ with discretized distance $d$, denoted as $n_{f,d}$. Guessing the exact value is unaffordable, so we guess it approximately, meaning that we guess $m_{f,d} \in \{0\} \cup \{(1+\epsilon)^r \mid r \in [0, \lceil \log_{1+\epsilon} n \rceil]\}$, and we say the guess is correct when $\frac{m_{f,d}}{1+\epsilon} \le n_{f,d} \le m_{f,d}$. The number of possible choice for $m_{f,d}$ is $O(\lceil \log_{1+\epsilon} n \rceil) = O(\frac{\ln n}{\epsilon})$.
    \item $\forall \emptyset \subsetneq X \subseteq F$ and $f \not\in X$, the number of clients in $C_X$ matched to $f$, denoted as $n_{X,f}$. It is also guessed approximately as above.
\end{itemize}

We use the guessing part to find a solution satisfying all conditions below:
\begin{itemize}
    \item $\forall f \in F, d \in \{(1+\epsilon)^{\lceil \log_{1+\epsilon} d(c,f) \rceil} \mid c \in C_I\}$, the number of clients matched from $C_I$ to $f$ with discretized distance $d$ is bounded by $m_{f,d}$;
    \item $\forall \emptyset \subsetneq X \subseteq F$, define $t_X = \max(0,|C_X| - \sum_{f \not\in X} m_{c,f})$, then the top-$t$ nearest clients to $X$ in $C_X$ are matched to $X$. All other clients in $C_X$ are arbitrarily allocated to facilities in $F \backslash X$, such that the set $A_f$ allocated to $f \not\in X$ have size at most $m_{X,f}$. Clients in $A_f$ can be matched to $f$ or $X$.
    \item All clients are matched to some facility, and the capacity constraint for all facilities is satisfied.
\end{itemize}
For the second condition, as the number of facilities matched in $X$ is understated, we should allow facilities allocated to facilities outside $X$ to match $X$ to fulfill the capacity constraint. However, measuring their cost as their distance to $f$ is always safe.

As the optimal solution simultaneously satisfies all above conditions with careful allocation in the second condition, there exists a solution. The solution can be easily found by a maximum matching algorithm in polynomial time. 

\paragraph{Approximation Ratio} When we know all the information exactly, Lemma \ref{lem:info} shows that the final solution gives $1+O(\epsilon)$ approximation. When guessing approximately, all information is at most $(1+\epsilon)$ times the correct value. From sub-additivity, exceeding part affects the norm by at most $\epsilon$ fraction. Finally, the approximation ratio is still $(1+O(\epsilon))$.

\paragraph{Runtime} All except the guessing part can be done in polynomial time. For the guessing, we need to guess $O(k^2)$ numbers with $O(\frac{\ln n}{\epsilon})$ choices each, so the total runtime is $(\frac{\ln n}{\epsilon})^{O(k^2)} n^{O(1)}$, which is bounded by $\left(\frac{k}{\epsilon}\right)^{O(k^2)}\cdot \poly(n)$.

\section{Missing Proofs from Section~\ref{sec:top-cn-kC}}

\lemmaconcave*
\begin{proof}
    Let $\delta, \delta' \in \calS_n$, and $\beta \in [0, 1]$. Let $\alpha$ and $\alpha'$ be the vectors that define $\overline{\topell}(\delta)$ and $\overline{\topell}(\delta')$ respectively in Definition~\ref{def:top-l-occur}. Then $\beta \alpha + (1-\beta)\alpha'$ is a candidate solution for the LP in the definition of $\overline{\topell}(\beta \delta + (1-\beta)\delta')$. Therefore, $\overline{\topell}(\beta \delta + (1-\beta)\delta') \geq  \overline{L_1}(\beta\delta + (1-\beta)\delta') = \beta\cdot \overline{L_1}(\delta) + (1-\beta)\cdot\overline{L_1}(\delta') = \beta \cdot \overline{\topell}(\delta) + (1-\beta)\cdot\overline{\topell}(\delta')$.
\end{proof}

\lemmadominatingaddingtransitive*
\begin{proof}
            Let $\delta = \sum_{h = 1}^H \beta_h\delta^h$ and $\delta' = \sum_{h = 1}^H \beta_h\delta'^h$. Our goal is to prove $\delta' \preceq_\gamma \delta$.  We need to show that $\overline{\topell}(\delta')\leq \gamma\cdot\overline{\topell}(\delta)$, for any $\ell \in [0, n]$. Fix $\ell \in [0, n]$. 
            
            Then, we have some $0 \leq \alpha \leq \delta'$ with $|\alpha'|_1 = \ell$ and $\overline{L_1}(\alpha') = \overline{\topell}(\delta')$. Then, there are vectors $\alpha'^0, \alpha'^1, \cdots, \alpha'^H \in \R_{\geq 0}^{\Z_{\geq 0}}$ satisfying
            \begin{itemize}
                \item $\alpha' = \sum_{h = 0}^H \beta_h \alpha'^h$, and
                \item $0 \leq \alpha'^h \leq \delta'^h$ for every $h \in [H]$.
            \end{itemize}
            
            For every $h \in [H]$, define $\ell_h = |\alpha'^h|_1$; then, $\overline{\topnorm{\ell_h}}(\delta'^h) \geq \overline{L_1}(\alpha'^h)$. There exists some $0 \leq \alpha^h \leq \delta^h$ such that $|\alpha^h|_1 = \ell_h$  and $\overline{\topnorm{\ell_h}}(\delta^h) = \overline{L_1}(\alpha^h)$. By that $\delta'^h \preceq_\gamma \delta^h$, we have 
            \begin{align*}
                \overline{L_1}(\alpha'^h)\leq \overline{\topnorm{\ell_h}}(\delta'^h) \leq \gamma \cdot \overline{\topnorm{\ell_h}}(\delta^h) = \gamma\cdot\overline{L_1}(\alpha^h).
            \end{align*}
            Also notice that $\sum_{h = 1}^H \beta_h\ell_h = \ell$. Define $\alpha = \sum_{h = 1}^H \beta_h\alpha^h$. Then, we have $0 \leq \alpha \leq \delta$, $|\alpha|_1 = \sum_{h=1}^H \beta_h|\alpha^h|_1 = \sum_{h=1}^H \beta_h\ell_h = \ell$, and 
            \begin{flalign*}
                &&\overline{L_1}(\alpha) &= \overline{L_1}\Big(\sum_{h = 1}^H \beta_h\alpha^h\Big) =\sum_{h = 1}^H \beta_h\overline{L_1}(\alpha^h)&&\\
                && &\geq \frac1\gamma\sum_{h=1}^H \beta_h \overline{L_1}(\alpha'^h) = \frac1\gamma\cdot\overline{L_1}\Big(\sum_{h = 1}^H \beta_h\alpha'^h\Big)=\frac1\gamma\cdot\overline{L_1}(\alpha') = \frac1\gamma\cdot \overline{\topell}(\delta'). &&\qedhere
            \end{flalign*} 
        \end{proof}

        \corodominatingaveraging*
        \begin{proof}
            We define $\delta^h$ for every integer $h \in [0, H]$ to be $n \cdot \frac{\theta^h}{|\theta^h|_1}$. Define $\delta'^0 = \delta^0$, and $\delta'^h = n \cdot {\bf e}_{b_h}$ for every $h \in [H]$. Then $\delta'^h \preceq_\gamma \delta^h$ for every integer $h \in [0, H]$; notice that $\gamma \geq 1$. Define $\beta_h = \frac{|\theta^h|_1}{n}$ for every $h \in [0, H]$. Then,
            \begin{align*}
                \delta = \sum_{h = 0}^H \beta_h\delta^h \qquad \text{and}\qquad \delta' = \sum_{h = 0}^H \beta_h\delta'^h.
            \end{align*}
            By Lemma \ref{lemma:dominating-adding-transitive}, we have $\delta' \preceq_\gamma \delta$.
        \end{proof}

        \thmmixoccurrence*
        \begin{proof}
        % [Proof of Theorem~\ref{thm:mix-occurrence}]
			There are two reals $x, y \in [0, 1]$ with $x \leq c \leq y$ such that
			\begin{align*}
				(1 - \alpha)x + \alpha y &= c, \quad\text{and}\quad\\
                \overline{\topcn}\big((1-\alpha)\delta + \alpha\cdot(\delta\oplus\delta'\otimes2)\big)  &= (1-\alpha)\cdot \overline{\topnorm{xn}}(\delta) + \alpha\cdot\overline{\topnorm{yn}}(\delta\oplus\delta'\otimes2).
			\end{align*}
		
			We have 
			\begin{align*}
				\overline{\topnorm{yn}}\big((\delta\oplus\delta'\otimes2)\big) \leq \overline{\topnorm{yn}}(\delta) + 2\cdot \overline{\topnorm{yn}}(\delta') \leq (1+2\gamma) \cdot \overline{\topnorm{yn}}(\delta).
			\end{align*}
			The second inequality is by that $\delta' \preceq_\gamma \delta$. Therefore,
            \begin{align*}
                \overline\topcn\big((1-\alpha) \delta + \alpha(\delta\oplus\delta'\otimes2)\big) &\leq (1-\alpha)\cdot \overline{\topnorm{xn}}(\delta) + \alpha \cdot \overline{\topnorm{yn}}(\delta) + 2\alpha\gamma \cdot \overline{\topnorm{yn}}(\delta)\\
                &\leq \overline{\topcn}(\delta) + 2\alpha\gamma \cdot \frac{1}{c} \cdot \overline{\topnorm{cn}}(\delta) \\
                &= \left(1 + \frac{2\alpha\gamma}{c}\right)\cdot \overline{\topnorm{cn}}(\delta).
            \end{align*}
			The second inequality used that $\overline{\topnorm{zn}}(\delta)$ for a fixed $\delta$ is a concave function over $z \in [0, 1]$, and $\overline{\topnorm{yn}}(\delta) \leq \overline{\topnorm{n}}(\delta) \leq \frac{1}{c} \cdot \overline{\topnorm{cn}}(\delta)$.
	\end{proof}	

% \section{Sampling using costs}
% \begin{theorem}
%     Let $v \in \R_{\geq 0}^n$ and $f:\R_{\geq 0}^n \to \R$ be a monotone symmetric norm. Let $\Delta \geq 1$ and $\epsilon \in (0, 1)$ be small enough. There is an efficient random algorithm that outputs a $S \gets [n]$ of size at most $O(\Delta \frac{\log n}{\epsilon})$ such that for every $T \subseteq [n]$ with $f(v[T]) \geq \frac{f(v)}{\Delta}$, we have $\Pr[S \cap T \neq \emptyset] \geq 1 - \frac1{n^3}$.
% \end{theorem}

% For every $\ell = \ceil{(1+\epsilon)^t} \in [n]$ for some integer $t$, we sample a set of elements from $[n]$. Consider the threshold for $\ell$. We sample 
    \section{$(3, 1+\frac2{ec} + \epsilon)$-Bi-Criteria Approximation for $(L_\infty, \topcn)$-Norms $k$-Clustering Problem}
\label{sec:2-norms}

In this section, we prove Theorem~\ref{thm:bi-criteria} by giving the $(3, 1 + \frac{2}{ec})$-bi-criteria approximation for the $(L_\infty, \topcn)$-norms $k$-clustering problem, for $c \in (\frac1e, 1]$. 

\begin{definition}\label{def: k-Clustering under (f, g)-Norm}
    In the $(L_\infty, \topcn)$-Norms $k$-Clustering problem,  we are given $F, m, C, n, k$ and $d$ as in Definition \ref{def: $k$-Clustering under $f$-Norm}, and a number $L \in \R_{\geq 0}$, the goal of the problem is to find a set $S \subseteq F$ of $k$ facilities so as minimize $|\vec d(S)|_\topcn$, subject to $|\vec d(S)|_\infty \leq L$. 

    Let $\opt$ be the optimum value of the instance. A solution $S \subseteq F, |S| = k$ is called an $(\alpha, \beta)$-bicriteria approximation for the instance, for some $\alpha, \beta \geq 1$, if $|\vec d(S)|_\infty \leq \alpha L$ and $|\vec d(S)|_\topcn \leq \beta\cdot\opt$.
\end{definition}

% \begin{definition}[Bicriteria-Approximation for $k$-Clustering under $(f, g)$-Norm]
%     Let an instance of the $(f, g)$-norms $k$-clustering problem be defined by $f, c, F$, $m, C, n, k, d$ and $L$ as in Definition \ref{def: k-Clustering under (f, g)-Norm}.  % \end{definition}

%So, in the setting of Theorem~\ref{thm:bi-criteria}, we are given an upper bound $L$ on $L_\infty$-norm of the clustering, where $L_\infty$ is a monotone symmetric norm. Our goal is to find a solution whose $\topcn$ cost is at most $(1+\frac2{ec}+\epsilon)\opt$, subject to the constraint that its $L_\infty$-norm objective is at most $(3+\epsilon)L$.  In particular, for the ($k$-center, $k$-median) objective, we achieve a tight approximation ratio of $(3, 1+\frac2e)$, improving upon the $(8, 4)$ approximation ratio due to Alamdari and Shmoys cite{Alamdari2017} achieved by a polynomial-time algorithm. \medskip

%TODO Need to be more formal. 

To prove Theorem~\ref{thm:bi-criteria}, the main modification to Algorithm~\ref{alg:choose-R-top-cn} is that in Step~\ref{step:Tpcn-calS}, after applying Theorem~\ref{thm:pseudo-approx} for the $\topcn$ norm to obtain $\calS$ and $S$, we add $O(k \log n)$ facilities that form a $1$-approximation (this can be achieved using LP rounding or Theorem~\ref{thm:pseudo-approx}) for the $L_\infty$ norm to $S$. Now $S$ is a $(1+O(\epsilon))$-approximation for $\topcn$ norm and has $L_\infty$ cost $L$. Then, when guessing $r_p$'s in Step~\ref{step:Tpcn-r}, we guarantee $r_p \leq L$.  In the end, we return the solution with the smallest $\topcn$ cost, whose $L_\infty$ cost at most $3L$. 

Clearly, if our guesses are correct, then the solution constructed have $L_\infty$ cost at most $3L$.  Consider an optimum cluster $(i^*, J)$, where every $j \in J$ has $d(i^*, j) \leq L$. If it is of type-1 or 2, the pivot $p$ is in $J$ and thus has $d(i^*, p) \leq L$. We are guaranteed to open a facility with distance $r_p \leq L$ to $p$.  Therefore, the distance of the facility to $i^*$ is at most $2L$, implying that the distance from all clients $J$ to $i^*$ is at most $3L$. When $(i^*, J)$ is of type-3, then the nearest facility to $i^*$ in $S$ has distance at most $2L$ to $i^*$, and thus distance at most $3L$ to all clients in $J$. Finally, for the case $c \in (0, \frac1e]$, a $(3, 3 + \epsilon)$-approximation can be achieved easily by modifying Algorithm~\ref{alg:pseudo} in the proof of Theorem~\ref{thm:pseudo-approx}. 
\end{document}